\documentclass[sort&compress
  ,final            
  ]
  {aipproc}

\usepackage{epsf}

\layoutstyle{6x9}

\usepackage{epsf}

\def\be{\begin{equation}}
\def\ee{\end{equation}}

\def\rtr{r_{\rm tr}}
\def\rign{r_{\rm ign}}

\def\rms{r_{\rm ms}}
	\def\dM{\dot{M}}
\def\Macc{M_{\rm acc}}
\def\Jacc{J_{\rm acc}}
\def\Jtot{J_{\rm tot}}
\def\Md{m}
\def\Jd{J}
\def\vK{v_{\rm K}}
\def\lK{l_{\rm K}}
\def\OmK{\Omega_{\rm K}}
\def\dMign{\dot{M}_{\rm ign}}
\def\dMop{\dot{M}_{\rm opaque}}
\def\dMtr{\dot{M}_{\rm trap}}
\def\dMM{\dot{M}_{0.1}}
\def\Kign{K_{\rm ign}}
\def\Kop{K_{\rm opaque}}
\def\Ktr{K_{\rm trap}}
\def\tvisc{t_{\rm visc}}
\def\tw{t_{\rm weak}}
\def\tinf{t_{\rm infall}}
\def\tacc{t_{\rm acc}}
\def\linf{l_{\rm infall}}
\def\rfr{R_\star}
\def\rc{r_{\rm circ}}
\def\rp{r_{\rm peak}}

\def\dq{\dot{q}_{\nu{\bar\nu}}}
\def\dE{\dot{E}_{\nu{\bar\nu}}}

\def\Teff{T_{\rm eff}}
\def\Eav{E_{\rm av}}

\newbox\grsign \setbox\grsign=\hbox{$>$} \newdimen\grdimen \grdimen=\ht\grsign
\newbox\simlessbox \newbox\simgreatbox \newbox\simpropbox
\setbox\simgreatbox=\hbox{\raise.5ex\hbox{$>$}\llap
     {\lower.5ex\hbox{$\sim$}}}\ht1=\grdimen\dp1=0pt
\setbox\simlessbox=\hbox{\raise.5ex\hbox{$<$}\llap
     {\lower.5ex\hbox{$\sim$}}}\ht2=\grdimen\dp2=0pt
\setbox\simpropbox=\hbox{\raise.5ex\hbox{$\propto$}\llap
     {\lower.5ex\hbox{$\sim$}}}\ht2=\grdimen\dp2=0pt
\def\simgt{\mathrel{\copy\simgreatbox}}
\def\simlt{\mathrel{\copy\simlessbox}}

\begin{document}

\title{Hyper-accreting black holes}

\classification{ }
\keywords{Accretion -- accretion disks -- dense matter -- gamma rays: bursts}

\author{Andrei M. Beloborodov}{
  address={Physics Department and Columbia Astrophysics Laboratory,
Columbia University, 538  West 120th Street New York, NY 10027, USA}
  ,altaddress={Also at Astro-Space Center of Lebedev Physical
Institute, Profsojuznaja 84/32, Moscow 117810, Russia} 
}

\begin{abstract}
Hyper-accretion disks are short-lived, powerful sources of neutrinos
and magnetized jets. Such disks are plausible sources of gamma-ray bursts. 
This review describes the disk structure, the neutrino conversion to 
electron-positron plasma around the disk, and the post-burst evolution.
\end{abstract}

\maketitle


\section{Introduction}

Hyper-accretion disks form when a neutron star merges with another compact 
object, neutron star or a black hole. Recent numerical simulations of 
mergers \cite{Shibata1,Etienne,Liu1} are fully relativistic and show how  
most of the mass of the binary system disappears behind the event horizon 
in about 10~ms, leaving a rotating debris disk around the black hole.
The mass of this centrifugally supported disk is $\Md\sim 0.01-0.1$\,M$_\odot$, 
similar to what was found in previous non-relativistic simulations 
(e.g. \cite{Ruffert,Rosswog}). The ensuing disk accretion is not followed 
by the merger simulations. It is established on a viscous timescale 
$\tvisc\sim 0.1$~s as the inner parts of the disk relax to a quasi-steady 
state; this relaxation was studied numerically in \cite{Setiawan,Shibata2}.
Most but not all of the disk is accreted on the timescale 
$\tvisc$, releasing an energy comparable to $\Md c^2$ and emitting copious 
neutrinos. This disk has an accretion rate
$\dM\sim ({\rm M}_\odot/s)\;(\Md/0.1\,{\rm M}_\odot)\;(\tvisc/0.1{\rm ~s})^{-1}$.

Similar neutrino-emitting disks may form during the core-collapse of 
massive stars if the stellar material has a sufficient angular momentum
\cite{Woosley,MacFadyen}. These hypothetical objects are often called 
``collapsars.'' After the formation of a central black hole of mass 
$M\sim 3$\,M$_\odot$ collapsars develop an accretion disk that 
is fed by the continually infalling stellar material.
The high accretion rate $\dM\sim 0.1$\,M$_\odot$~s$^{-1}$ is sustained 
for $\sim 10$~s (the core-collapse timescale). 
Recent relativistic MHD simulations of this accretion show 
how the black hole could accumulate a large magnetic flux and create 
jets via the Blandford-Znajek process \cite{Barkov}.

The studies of hyper-accretion disks are greatly stimulated by observations 
of cosmological gamma-ray bursts (GRBs, see \cite{Piran} for a review). 
Hyper-accretion is expected to produce hyper-jets on a timescale 
$\sim 0.1-10$~s. If a fraction $\epsilon_{\rm jet}$ of the accretion power 
$\dot{M}c^2$ is channeled to a relativistic jet, it leads to an explosion 
with energy $E_{\rm jet}\sim 2\times 10^{51}(M_{\rm acc}/{\rm M}_\odot)
 (\epsilon_{\rm jet}/10^{-3})$~erg, 
where $M_{\rm acc}$ is the mass accreted through the disk.  
The energy and duration of the jet is consistent with GRB observations.

Hyper-accretion disks are markedly different from normal accretion 
disks in X-ray binaries and AGN. Their optical depth to photon scattering 
is enormous and radiation is trapped inside the disk, being advected by 
the matter into the black hole.
However, the disk can be efficiently cooled by neutrino emission. 
Significant neutrino losses can occur when $\dot{M}>10^{-3}$\,M$_\odot$~s$^{-1}$ 
and make the disk relatively thin and neutron rich.

The accreting black hole is expected to have a significant angular 
momentum, because it forms from rotating matter and is further spun up by 
accretion. The black-hole spin helps the jet formation through the 
Blandford-Znajek process. It also affects the black-hole spacetime in 
such a way that the disk extends to smaller radii and 
the overall efficiency of accretion significantly increases. 
For example, the inner radius of the disk
around a maximally-rotating black hole (spin parameter $a=1$) 
is reduced by a factor of 6 compared with the Schwarzschild case $a=0$.
This leads to a higher temperature and a higher neutrino intensity
above the disk, increasing the rate of neutrino annihilation into $e^\pm$
pairs. 
Therefore, disks around rapidly spinning black holes can deposit an 
interesting fraction of their energy into the $e^\pm$ plasma outside the
disk and facilitate the formation of ultra-relativistic jets.

The size of a hyper-accretion disk depends on the specific angular momentum 
of the accreting matter, $l$, which is modest in neutron-star mergers 
and probably even smaller in collapsars. 
The accretion flows in collapsars are quasi-spherical and
may form a special ``mini-disk'' that is not supported centrifugally and 
instead accretes on the free-fall time.
Larger $l$ leads to standard viscous accretion, 
which leaves a relict disk carrying the initial angular momentum of the 
accreted matter. The relict disk gradually spreads to larger radii, and
its late evolution may be relevant to the post-burst activity of GRBs.


\section{Viscous disks}

As matter spirals into the black hole, it is viscously heated:
the gravitational energy is converted to heat.
The heat is distributed between nuclear matter, radiation, and $e^\pm$ pairs,
in perfect thermodynamic equilibrium.
In particular, the equilibrium $e^\pm$ population is maintained.
As discussed below, electrons are mildly degenerate in neutrino-cooled
disks, which affects the $e^\pm$ density.
The equilibrium microphysics is determined by only three parameters: 
temperature $T$, baryon mass density $\rho$, and electron fraction $Y_e$ 
(equal to the charged nucleon fraction). Other parameters ---
e.g. the electron chemical potential $\mu_e$ and density of $e^\pm$ pairs, 
$n_\pm$ --- are derived from $T$, $\rho$ and $Y_e$.
At radii $r\simlt 10^8$~cm, temperature and density are high enough
to maintain the nuclear statistical equilibrium, which determines the 
abundances of all nuclei. Nuclear matter in the disk is dissociated into 
free nucleons $n$ and $p$ inside radius $r_\alpha=(40-100)r_g$ where 
$r_g\equiv 2GM/c^2$. The temperature at this radius is $kT\simlt 1$~MeV.

Neutrino cooling is significant in the inner region where $kT>1$~MeV.
The far dominant mechanism of neutrino emission is the $e^\pm$
capture onto nucleons:  
\begin{equation}
\label{eq:reactions}
  e^-+p\rightarrow n+\nu_e, \qquad e^++n\rightarrow p+\bar{\nu}_e.
\end{equation}
The escaping neutrinos not only cool down the disk --- they also change
its electron fraction $Y_e$ if the emission rates of $\nu$ and $\bar\nu$ 
are not exactly equal. The first goal of the disk modeling is to find $T$, 
$\rho$, $Y_e$, and self-consistently evaluate the neutrino losses.

\subsection{Modeling disk accretion}

Accretion is quasi-steady in the inner region of 
the disk where $\tvisc$ is shorter than the timescale of $\dM$ evolution. 
Viscosity in accretion disks is caused by MHD turbulence, 
sustained by the magneto-rotational instability \cite{Balbus}.
It creates an effective kinematic viscosity coefficient $\nu$, which may
be related to the half-thickness of the disk, $H$, and sound speed, $c_s$:
~$\nu=\alpha c_sH$, where $\alpha\sim 0.01-0.1$ is a dimensionless parameter. 
A number of works studied hyper-accretion disks with this traditional 
parameterization of viscosity \cite{Popham,Narayan1,diMatt,Kohri,CB,Kawanaka,Janiuk}.
Two works constructed models in full general relativity and studied
disks around spinning black holes \cite{Popham,CB}. 
The following list highlights advances and limitations of the
current models \cite{CB}:

\begin{list}{--}{\setlength{\labelwidth}{4mm} \setlength{\leftmargin}{5mm}}
\item The model is fully relativistic: the disk dynamics is calculated in
Kerr metric.


\item Hydrodynamic equations are solved with the vertically-integrated $\alpha$ 
prescription. The equations include radial transport of heat and lepton number.
 

\item Local microphysics is calculated exactly: nuclear composition, electron 
degeneracy, neutrino emissivity and opacity etc., using the equilibrium 
distribution functions for all species except neutrinos. Neutrinos are
modeled separately in the opaque and transparent zones of the disk,
matching at the transition between the two zones.


\item The model provides only vertically averaged $T$, $\rho$, and $Y_e$
(which are approximately equal to their values in the midplane of the disk).
The diffusion of neutrinos in the opaque zone is treated in the simplest 
one-zone approximation (using escape probability).
This gives a good approximation for the energy losses, however, does
not give the exact neutrino spectrum emerging from the opaque zone.

\end{list}

The vertically-integrated approximation 
provides no information about the vertical structure of the disk and 
its corona. The vertical structure may be eventually understood with
global 3D time-dependent MHD simulations that include energy losses, 
although the results of such simulations generally depend on the assumed 
initial magnetic configuration. The behavior of magnetic field on large 
scales is coupled to the local turbulence cascade that extends to scales 
$\ll H$. The microscopic magnetic Prandtl number for hyper-accretion 
disks has been recently estimated in \cite{Rossi1}. The models discussed 
here are aware of the MHD issues only through the value of $\alpha$. They 
are computationally much cheaper than MHD simulations and allow one to 
study the disk in a broad range of $\dM$ and $\alpha$.

Vertically-integrated disks are described by 1D equations that express 
conservation of
baryon number, energy, and momentum (angular and radial) in Kerr spacetime
(see \cite{B99} for a review). The full set of these equations can be solved 
\cite{B98}, and the solutions show that the deviation from circular Keplerian 
rotation is small ($\simlt 10$\%) even when the disk is strongly 
advective (i.e. when the released heat is transported radially without 
losses).\footnote{
    A strong reduction of $\Omega$ below $\Omega_{\rm K}$ can occur in the 
    limit of a large steady disk with no cooling. This limit does not apply 
    to hyper-accretion disks which are transient and have a moderate radius.} 
Thus, the angular velocity of the disk can be approximated by its Keplerian 
value $\OmK$. A small radial velocity is superimposed on this rotation:
$u^r=-\alpha\,S^{-1}\,c_s\,(H/r)$, where $c_s=(P/\rho)^{1/2}$ is the 
isothermal sound speed, $H$ is the half-thickness of the disk; $S(r)$ is
a numerical factor determined by the inner boundary condition \cite{CB}. 
This description of the velocity field in the disk is a good approximation 
everywhere except in the very vicinity of the inner boundary where $|u^r|$ 
exceeds $c_s$.

\medskip

\begin{figure}[tp]
\includegraphics[height=0.5\textwidth]{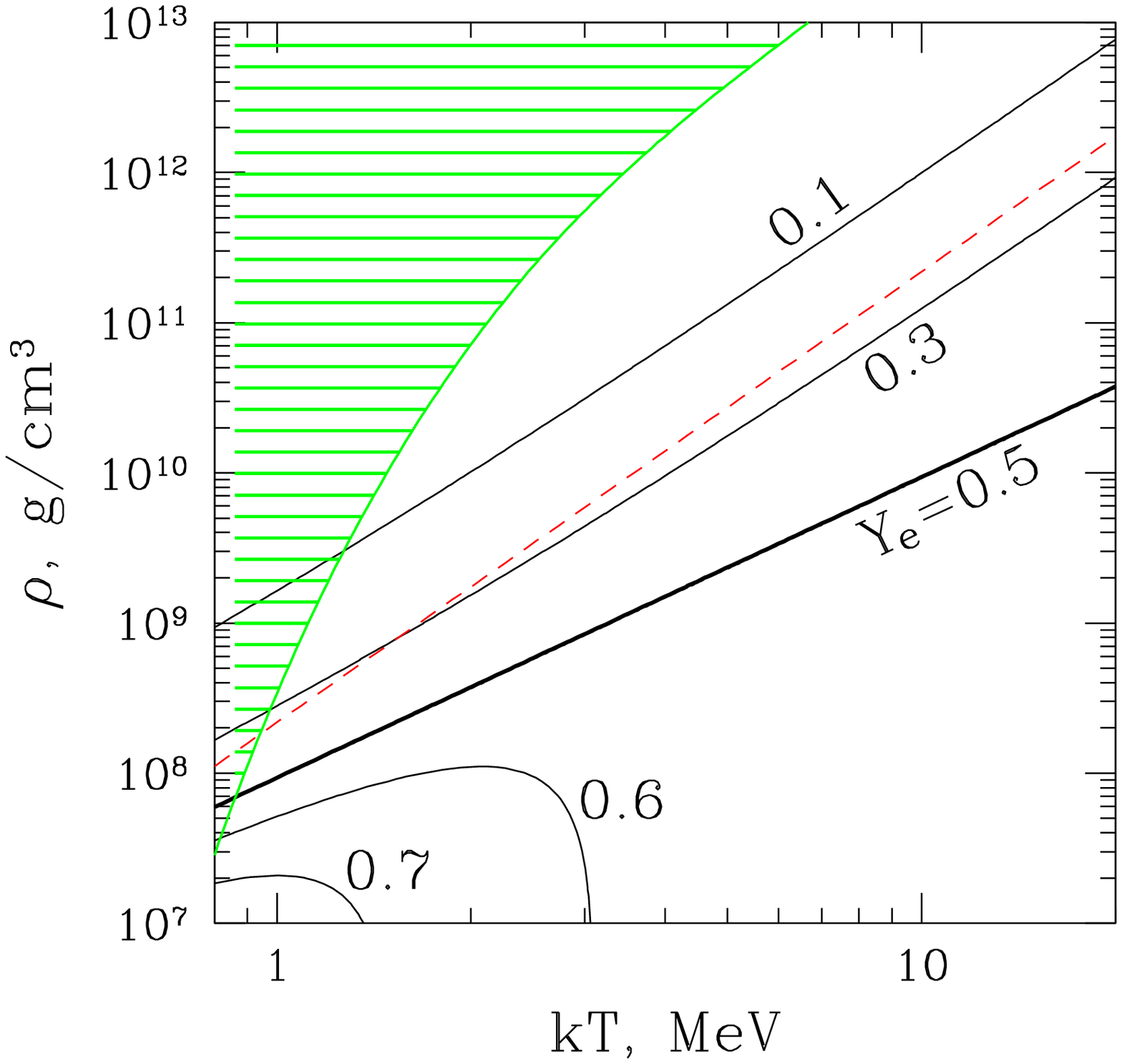} \hfill
\includegraphics[height=0.5\textwidth]{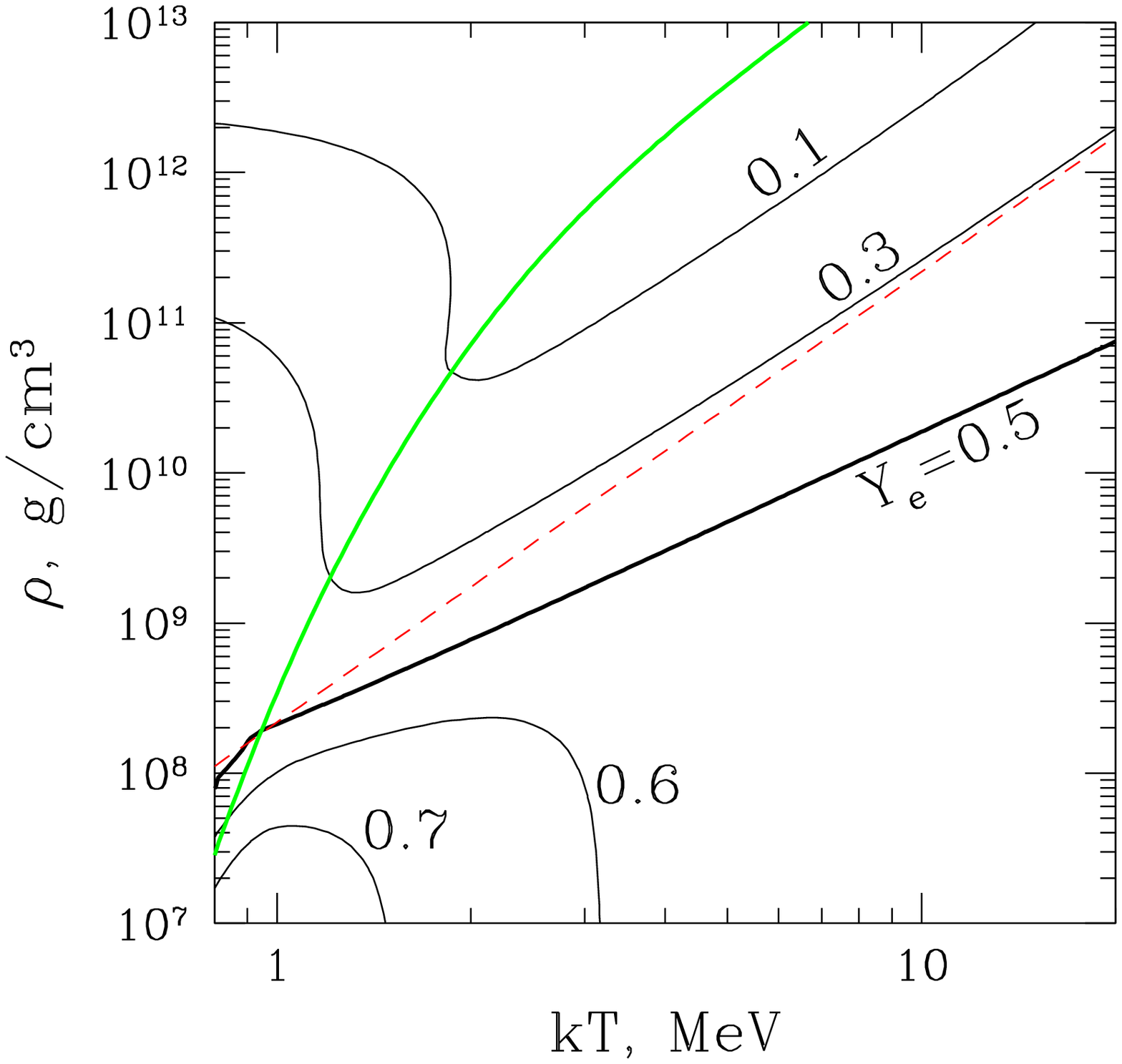}
\caption{{\bf Left panel}: Contours of the equilibrium $Y_e(T,\rho)$ on the 
$T$-$\rho$ plane for $\nu$-transparent matter. The electrons become
degenerate near the dashed line given by 
$kT_{\rm deg}=\hbar c(\rho/m_p)^{1/3}=7.7\rho_{11}^{1/3}$~MeV.
The $Y_e$ contours are calculated assuming that the nuclear matter is 
dissociated into free nucleons; they are invalid in the shaded region 
where matter is dominated by composite nuclei. The ``neutronization line'' 
$Y_e=0.5$ is given by $kT_n=33\rho_{11}^{1/2}$~MeV.
{\bf Right panel}: The equilibrium $Y_e(T,\rho)$ for 
$\nu$-opaque matter with neutrino chemical potential $\mu_\nu=0$.
The free-nucleon region is the same as in the left panel. 
The calculation of the equilibrium $Y_e$ is now extended 
into the region of composite nuclei. The neutronization line $Y_e=0.5$ 
is given by $kT_n=23.1\rho_{11}^{1/2}$~MeV. 
}
\end{figure}

In contrast to accretion disks in X-ray binaries and AGN, there is
one more conservation law that must be taken into account:
conservation of lepton number, 
\be
\label{eq:Ye}
    \frac{1}{H}(\dot N_{\bar\nu}-\dot N_{\nu})
     =u^r\left[ \frac{\rho}{m_p}\frac{dY_e}{dr}+\frac{d}{dr}
      (n_\nu -n_{\bar\nu}) \right].
\ee
Here $\dot N_{\nu}$ and $\dot N_{\bar\nu}$ are the number fluxes of
neutrinos and anti-neutrinos per unit area (from one face of the disk),
$n_\nu$ and $n_{\bar\nu}$ are the number densities of neutrinos and
anti-neutrinos inside the disk. This equation determines $Y_e$, which 
is related to the neutron-to-proton ratio by $Y_e=(n_n/n_p+1)^{-1}$ and 
greatly affects the rate of neutrino cooling. 

\medskip

In the models solved in \cite{CB} and shown below, $Y_e$ is calculated 
using Eq.~(\ref{eq:Ye}). Note, however, that throughout 
most of the {\it neutrino-cooled} disk, the right side of 
Eq.~(\ref{eq:Ye}) is small compared with each of the two terms on the 
left side, and $Y_e$ is nearly equal to the local equilibrium value such that 
$\dot N_{\bar\nu}\approx\dot N_{\nu}$. This equilibrium $Y_e$ is determined
by the local temperature and density and found for both neutrino-opaque and 
neutrino-transparent matter \cite{Imsh,Arnett,B03a}. It is shown in Fig.~1.


 \begin{figure}[htp]
  \includegraphics[height=0.5\textwidth]{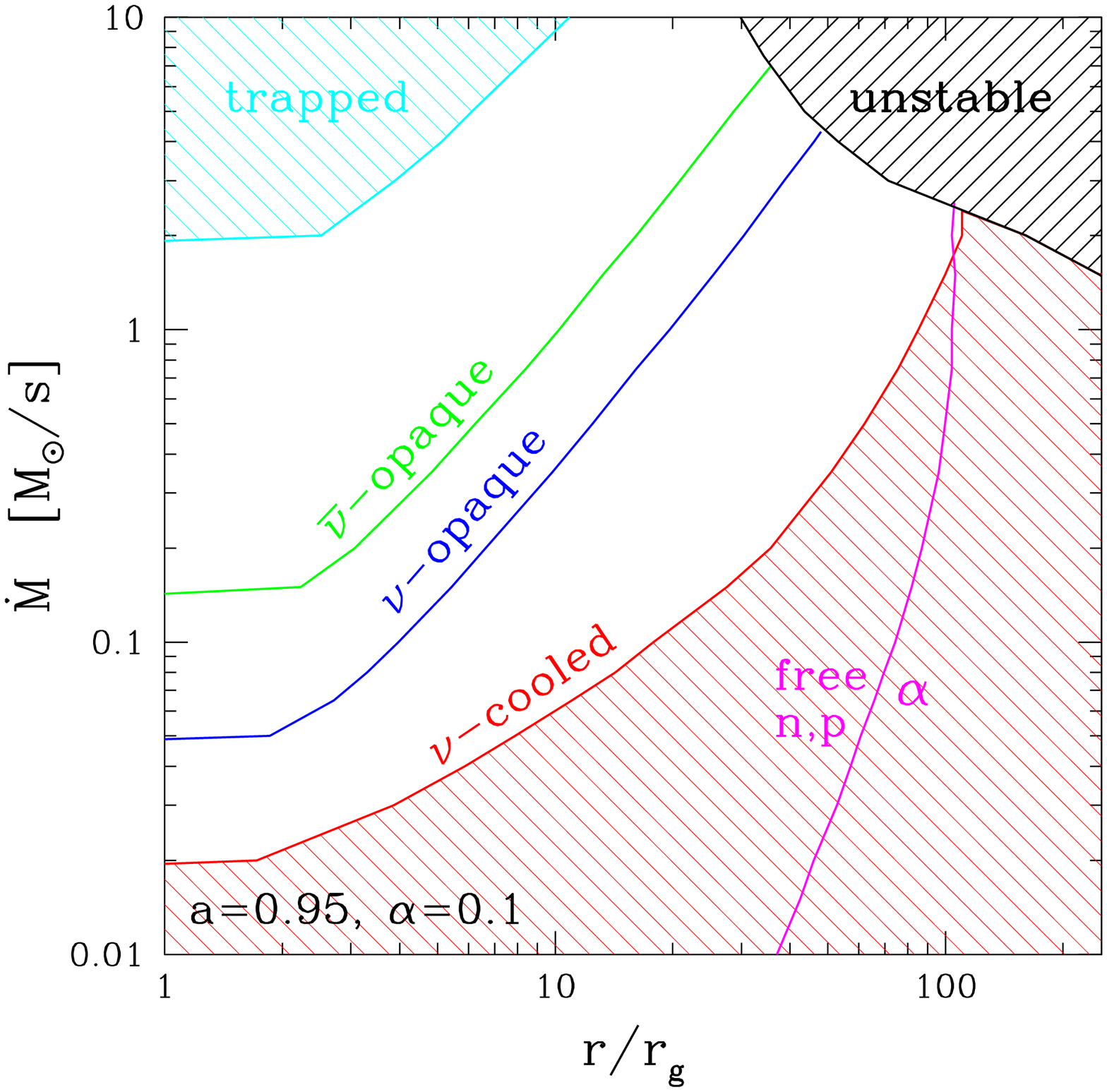} \hfill
  \includegraphics[height=0.5\textwidth]{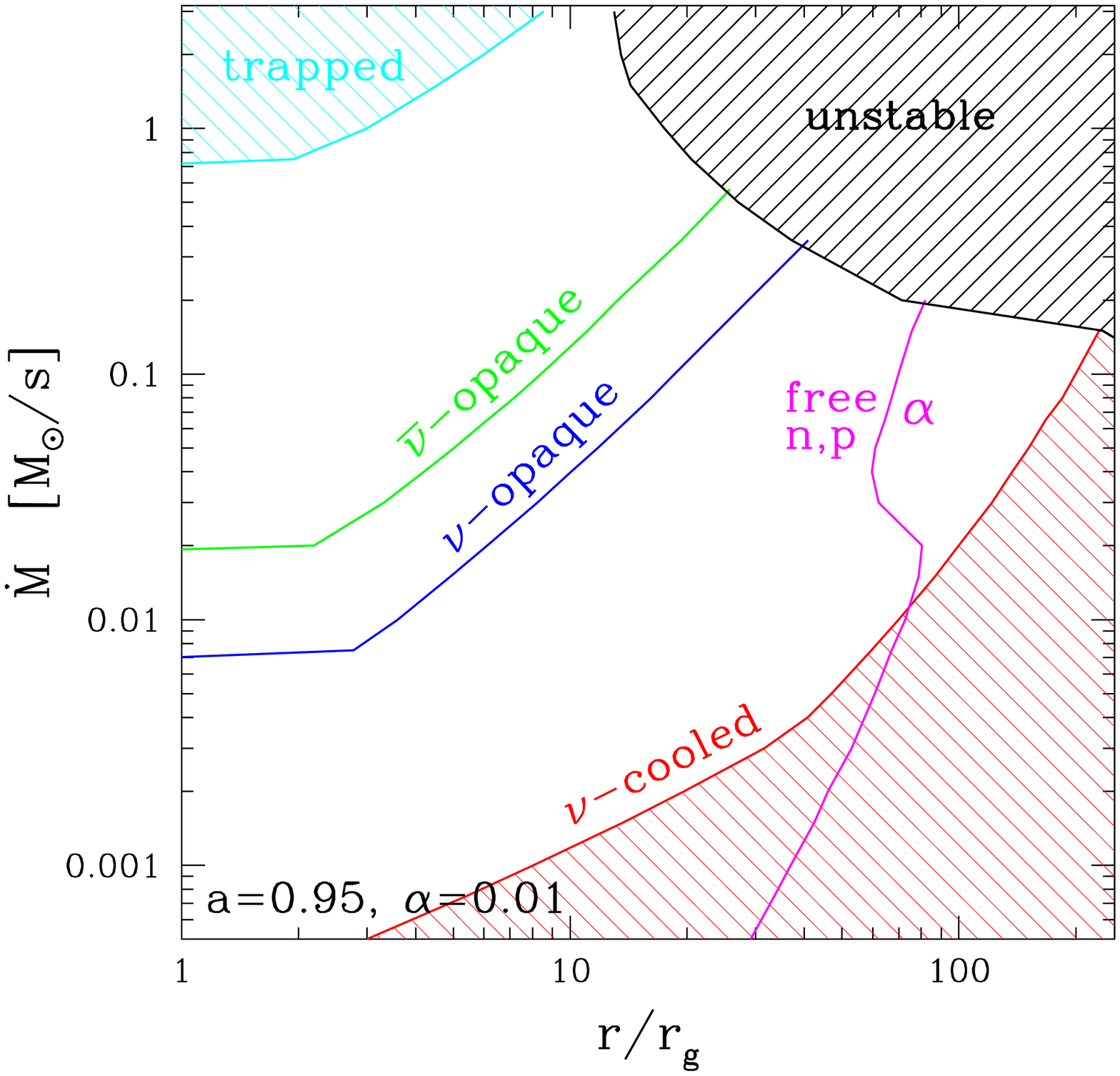}
  \caption{
Boundaries of different regions on the $r$-$\dM$ plane for disks around
a black hole of mass $M=3$\,M$_\odot$ and spin parameter $a=0.95$.
Neutrino cooling is inefficient in the shaded region below the "$\nu$-cooled"
curve and above the ``trapped'' curve.
The shaded region marked ``unstable'' is excluded: the steady model is
inconsistent in this region because of the gravitational instability.
The disk extends down to the marginally stable orbit of radius 
$\rms\approx r_g$ where $r_g=2GM/c^2$.
{\bf Left panel}: Disks with viscosity parameter $\alpha=0.1$.
{\bf Right panel}: Disks with viscosity parameter $\alpha=0.01$.
(From \cite{CB}.)
}
 \end{figure}

\subsection{Overview of disk properties}

Hyper-accretion disks have several zones separated by the 
following characteristic radii:

\begin{enumerate}{\setlength{\leftmargin}{5mm}}

\item Radius $r_\alpha$ where 50\% of $\alpha$-particles are decomposed
into free nucleons. The destruction of $\alpha$-particles consumes 7~MeV
per nucleon, which makes the disk thinner.

\item "Ignition" radius $\rign$ where neutrino emission switches on.
At this radius, the mean electron energy becomes comparable to
$(m_n-m_p)c^2$, enabling the capture reaction $e^-+p\rightarrow n+\nu$.
Then neutrino cooling due to reactions~(\ref{eq:reactions}) becomes
significant, further reducing the disk thickness $H/r$. 

\item Radius $r_\nu$ where the disk becomes opaque to neutrinos
and they relax to a thermal distribution. The disk is still
almost transparent to anti-neutrinos at this radius.

\item Radius $r_{\bar\nu}$ where the disk becomes opaque to anti-neutrinos,
so that both $\nu$ and ${\bar\nu}$ are now in thermal equilibrium with
the matter. The disk is still cooled efficiently at this
radius since $\nu$ and ${\bar\nu}$ diffuse and escape the flow faster
than it accretes into the black hole.

\item Radius $\rtr$ where neutrino diffusion out of the disk becomes slower 
than accretion, and neutrinos get trapped and advected into the black hole.

\end{enumerate}

\noindent 
The different zones of the disk are shown on the $\dot{M}-r$ diagram in 
Fig.~2. In addition, this figure shows the zone of gravitational instability.

Three characteristic accretion rates can be defined:
$\dMign$ above which the disk is neutrino-cooled in the inner region,
$\dMop$ above which the disk is opaque to neutrinos in the inner region,
and $\dMtr$ above which the trapping of neutrinos occurs in the inner region. 
The dependence of $\dMign$, $\dMop$, and $\dMtr$ on $\alpha$ is well
approximated by the following power laws \cite{CB},
\be
\label{eq:pw}
   \dMign=\Kign\left(\frac{\alpha}{0.1}\right)^{5/3}, \qquad
   \dMop=\Kop\left(\frac{\alpha}{0.1}\right), \qquad
   \dMtr=\Ktr\left(\frac{\alpha}{0.1}\right)^{1/3}.
\ee
The normalization factors $K$ depend on the black hole spin $a$.
For $a=0.95$ they are $\Kign=0.021$\,M$_\odot$~s$^{-1}$, 
$\Kop=0.06$\,M$_\odot$~s$^{-1}$, $\Ktr=1.8$\,M$_\odot$~s$^{-1}$
and for $a=0$ they are
$\Kign=0.071$\,M$_\odot$~s$^{-1}$, $\Kop=0.7$\,M$_\odot$~s$^{-1}$, 
$\Ktr=9.3$\,M$_\odot$~s$^{-1}$.


\medskip

To complete this short guide to quasi-steady viscous disks, Figs.~3-4 show 
$T$, $\rho$, $Y_e$, and $H/r$ for a model with 
$\dot{M}=0.2$\,M$_\odot$~s$^{-1}$, $M=3$\,M$_\odot$, and $a=0.95$.

\begin{figure}[htp]
\includegraphics[height=0.4\textwidth]{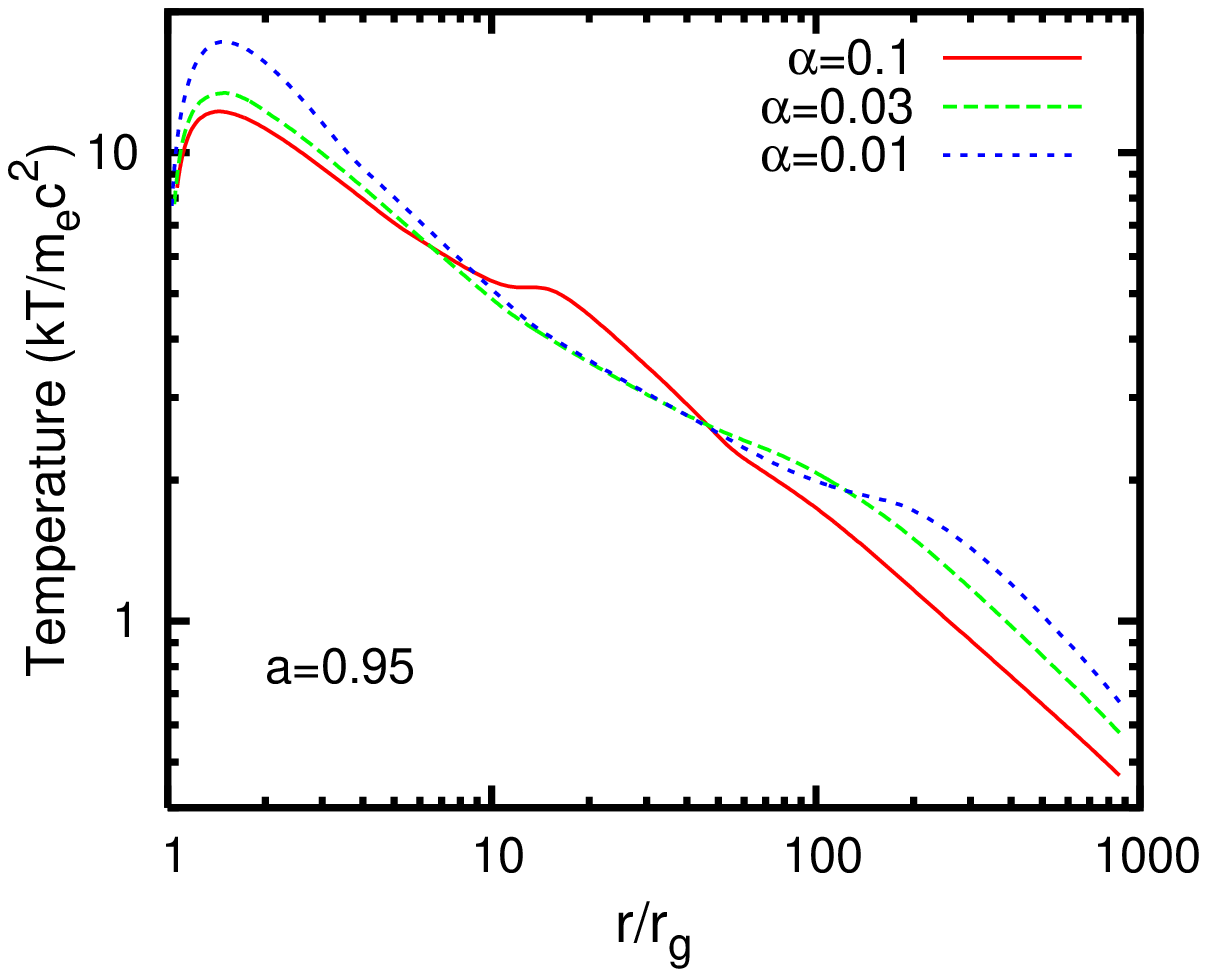} \hfill
\includegraphics[height=0.4\textwidth]{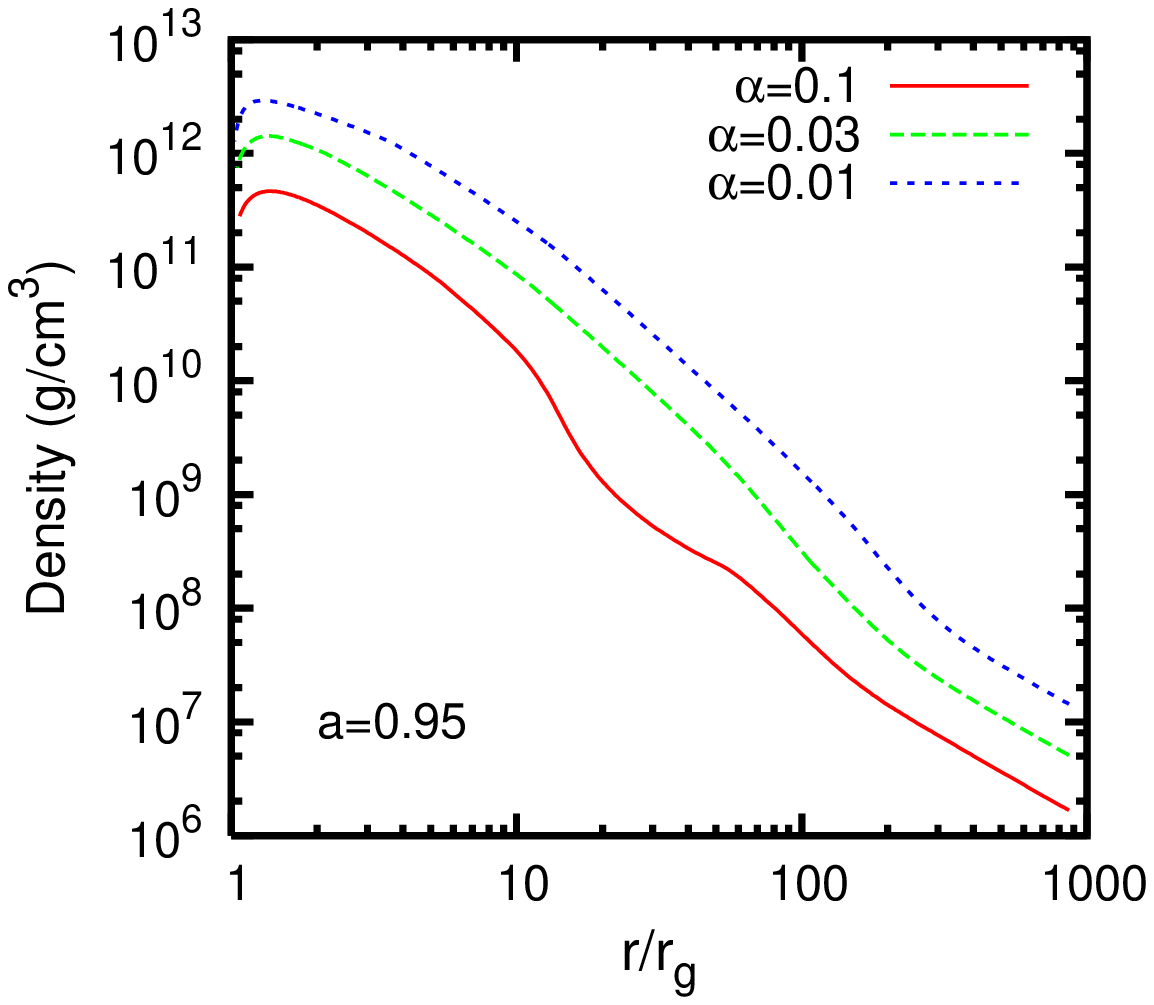}
\caption{
Disk with $\dM=0.2$\,M$_\odot$s$^{-1}$ around a black hole of
mass $M=3$\,M$_\odot$ and spin $a=0.95$. Three models are shown with 
viscosity $\alpha=0.1$, 0.03, and 0.01. Radius is measured in units
of $r_g=2GM/c^2=10$~km.
{\bf Left panel}: Temperature in units of $m_ec^2$.
{\bf Right panel:} Mass density.
(From \cite{CB}.)}
\end{figure}
\begin{figure}[htp]
\includegraphics[height=0.41\textwidth]{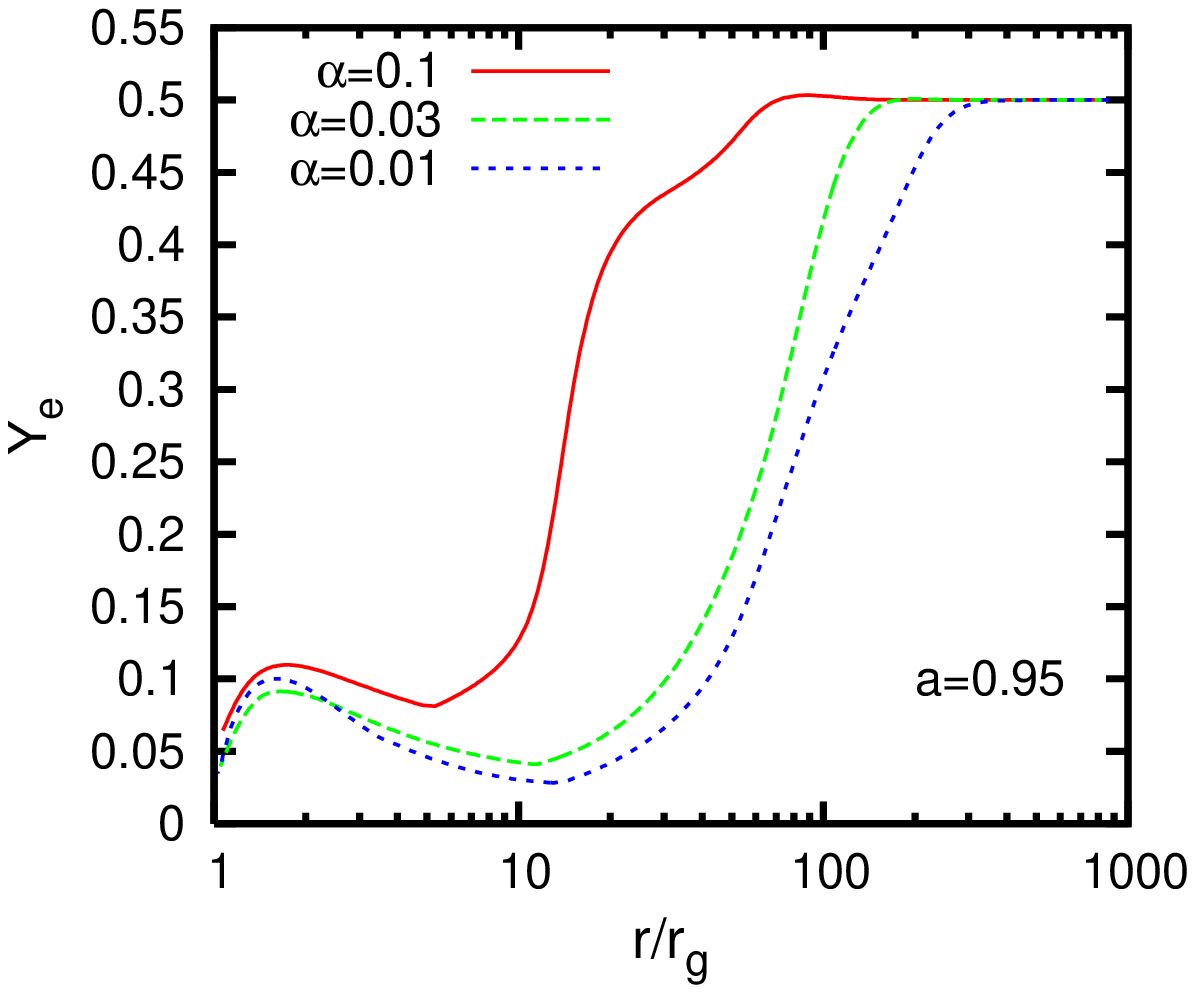} \hfill
\includegraphics[height=0.41\textwidth]{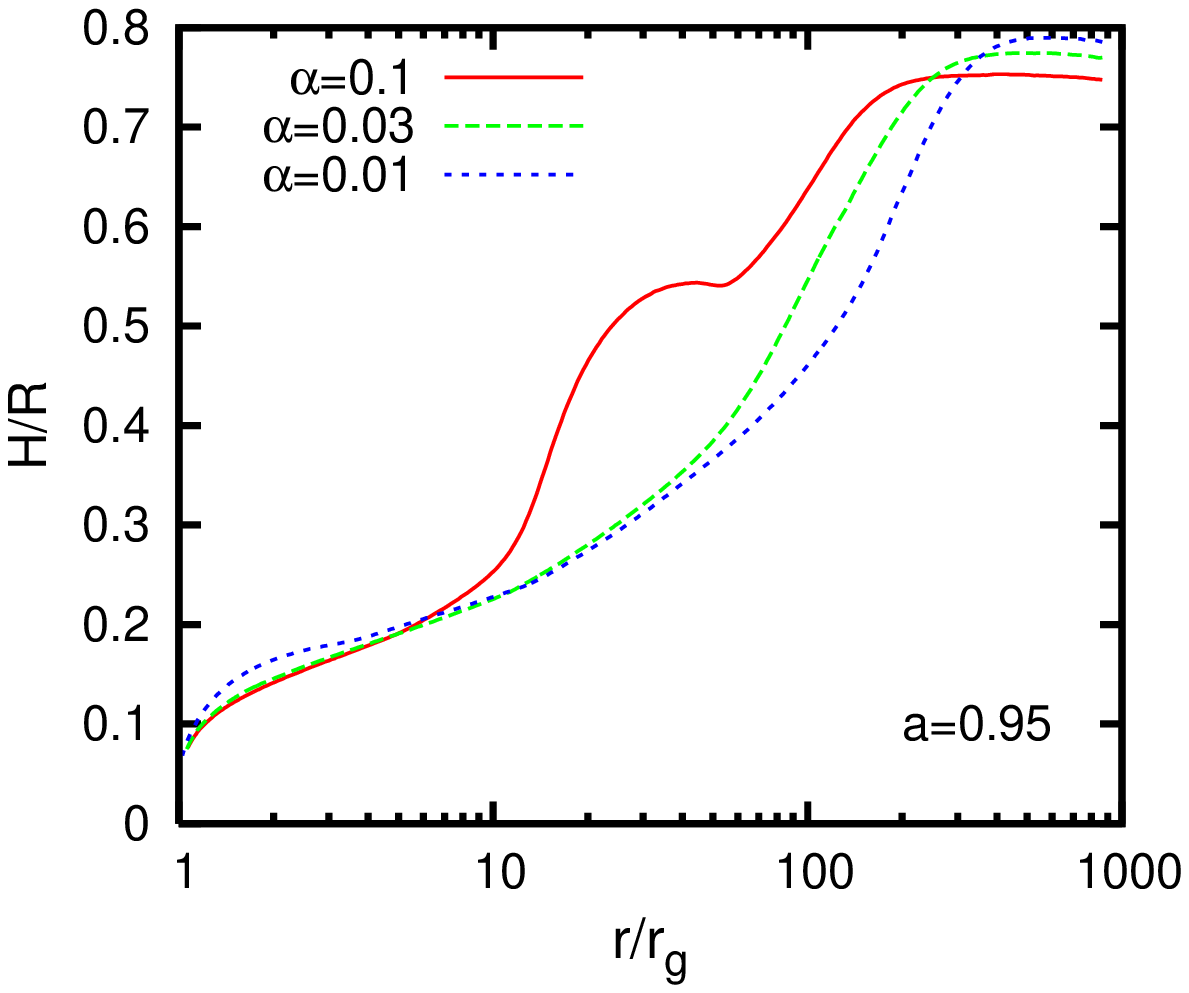}
\caption{Electron fraction $Y_e$ and thickness of the disk $H/r$ for 
the same three models as in Fig.~3.}
\end{figure}

The main properties of the {\it neutrino-cooled} disk (i.e. at $r<\rign$)
may be summarized as follows \cite{CB}.

\begin{list}{$\circ$}{\setlength{\leftmargin}{5mm}}

\item The disk is relatively thin, $H/r\sim 0.1-0.3$,
especially in the inner region where most of accretion energy is released.

\item The $\nu$-cooled disk is locally very close to $\beta$-equilibrium, 
$\dot{N}_\nu\approx\dot{N}_{\bar\nu}$. In particular, the relation between
$\rho$, $T$, and $Y_e$ calculated under the equilibrium assumption
(Fig.~1) is satisfied with a high accuracy.

\item Degeneracy of electrons in the disk significantly suppresses
the positron density $n_{e^+}$. However, the strong degeneracy limit is
not applicable --- the disk regulates itself to a mildly degenerate state
with $\mu_e/kT=1-3$.
The reason of this regulation is the negative feedback of degeneracy on
the cooling rate: higher degeneracy $\mu_e/kT$ $\rightarrow$ fewer
electrons (lower $Y_e$) and positrons ($n_{e^+}/n_{e^-}\sim e^{-\mu_e/kT}$) 
$\rightarrow$ weaker neutrino emission $\rightarrow$
lower cooling rate $\rightarrow$ higher temperature $\rightarrow$ lower
degeneracy.

\item Pressure in $\nu$-cooled disks is dominated by baryons,
$P\approx P_b=(\rho/m_p)kT$, most of which are neutrons (Fig.~5 shows
contributions to pressure for two sample models).

\item All $\nu$-cooled disks are very neutron rich in the inner region,
with $Y_e\sim 0.1$ or lower.

\end{list}

\begin{figure}[t]
\includegraphics[height=0.44\textwidth]{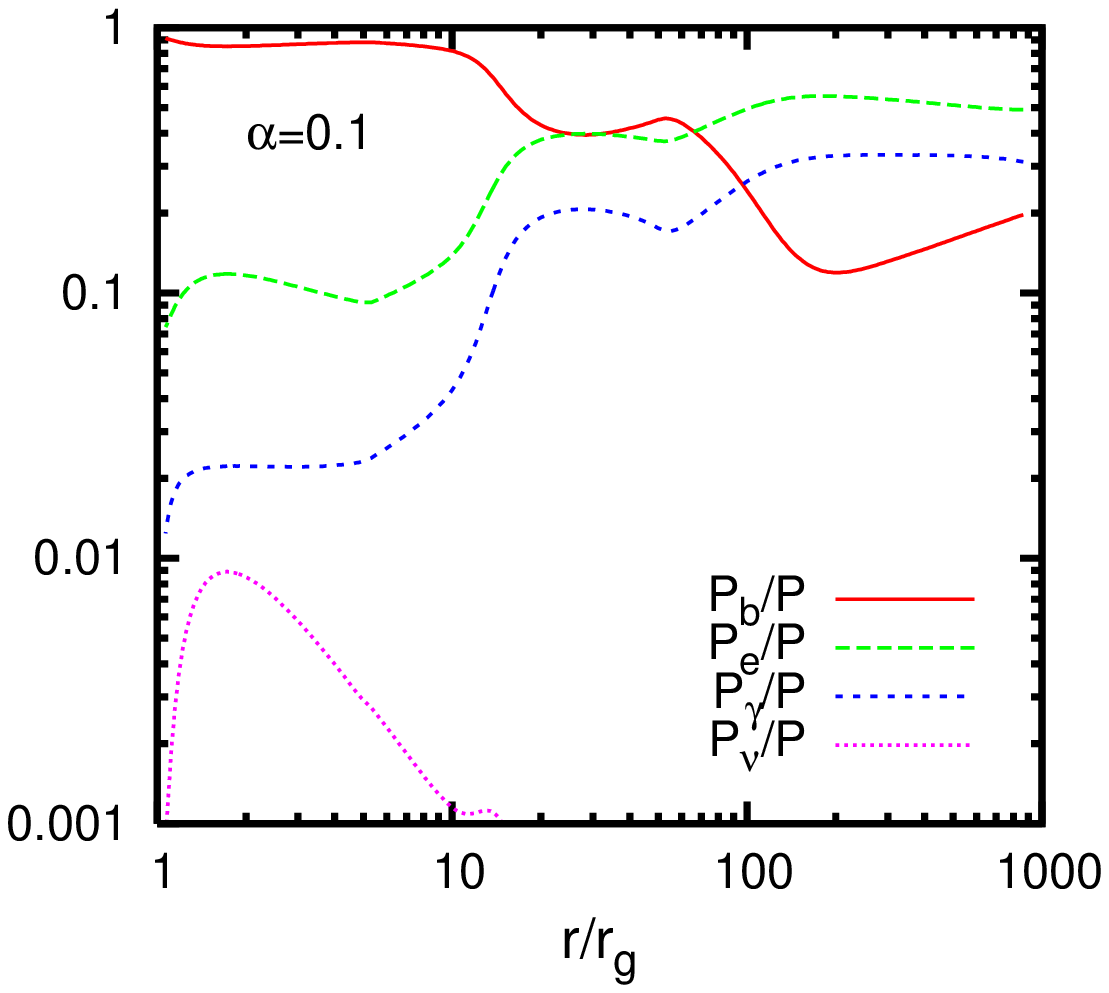} \hfill
\includegraphics[height=0.44\textwidth]{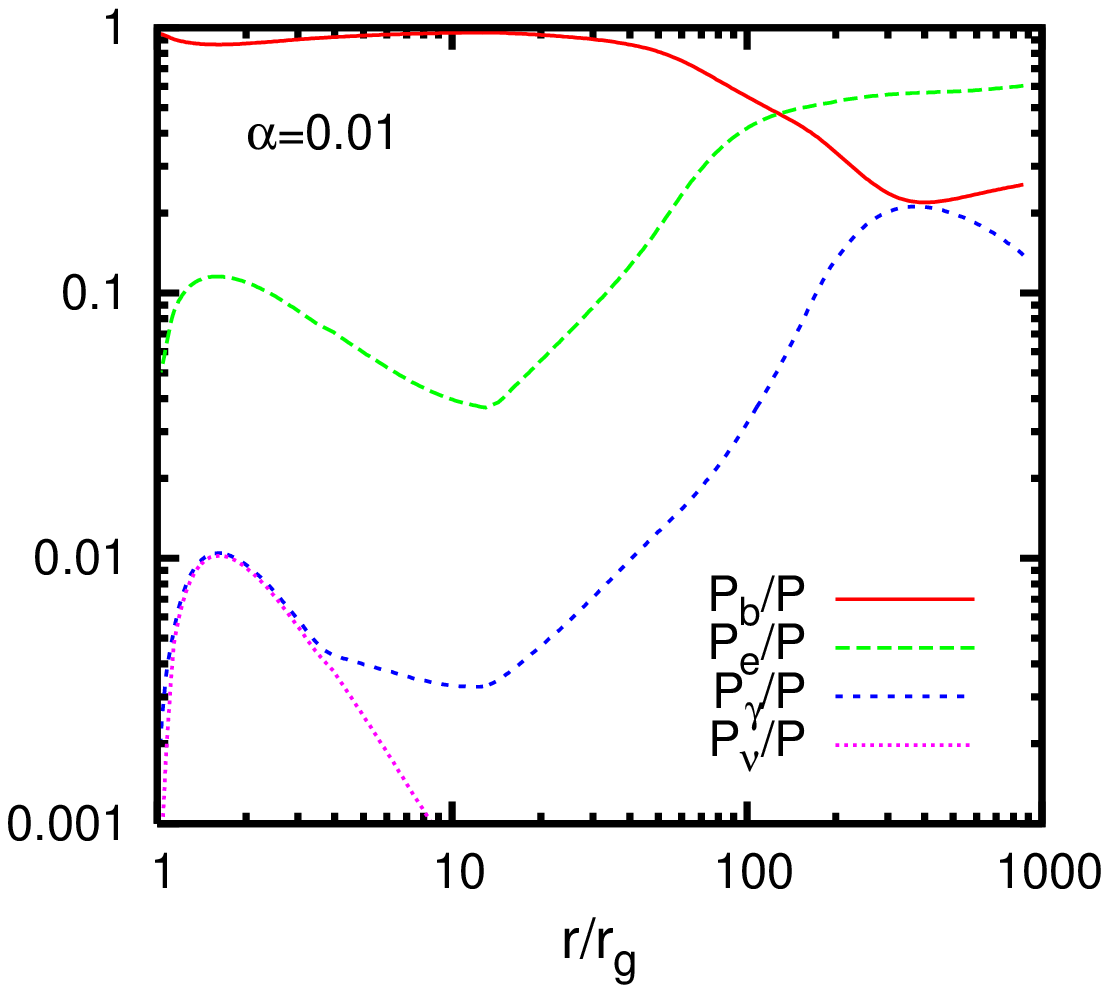}
\caption{
Contributions to total pressure $P$ from baryons,  $P_b$,
electrons and positrons $P_e=P_{e^-}+P_{e^+}$, radiation $P_\gamma$,
and neutrinos $P_\nu+P_{\bar\nu}$ for the 
accretion disk with $\dM=0.2$\,M$_\odot$~s$^{-1}$
around a black hole of mass $M=3$\,M$_\odot$ and spin $a=0.95$. 
{\bf Left panel:} Model with viscosity parameter $\alpha=0.1$. 
{\bf Right panel:} Model with $\alpha=0.01$. 
(From \cite{CB}.)}
\end{figure}


\section{Neutrino annihilation around the disk}

The emitted neutrinos and anti-neutrinos can collide and convert to $e^\pm$, 
thereby depositing energy \cite{Goodman}.
The emission of tau and muon neutrinos is negligible \cite{CB},
so only reaction $\nu_e +\bar\nu_e\rightarrow e^-+e^+$ can be considered.
Its cross section (assuming center-of-momentum energy $\gg m_ec^2$) 
is given by $\sigma_{\nu\bar\nu}\approx 3.3\times 10^{-45}
({\mathbf p}_\nu{\mathbf\cdot}{\mathbf p}_{\bar\nu})^2
(p_\nu^0 p_{\bar\nu}^0)^{-1}$cm$^2$
where ${\mathbf p}_\nu$, ${\mathbf p}_{\bar\nu}$ are the 4-momenta of $\nu_e$
and $\bar\nu_e$, expressed in units of $m_ec$. The cross section is small 
and only a small fraction $\epsilon\sim 10^{-3}-10^{-2}$ of the total 
neutrino luminosity $L$ converts to $e^\pm$ plasma. Nevertheless, this 
energy may be sufficient to drive a relativistic jet (or help the formation 
of a magnetically-dominated jet) since it occurs above the disk where the 
mass density is relatively low, especially near the rotation axis 
\cite{Eichler}.

\medskip

Neutrinos emitted by the disk follow null geodesics in Kerr spacetime.
The efficiency $\epsilon$ of their annihilation can be calculated 
numerically by tracing the geodesics, evaluating the local energy deposition 
rate $\dq$ [erg~s$^{-1}$~cm$^{-3}$] everywhere around the black hole,
and then integrating over volume to obtain the net energy deposition rate
$\dE$ (energy at infinity per unit time at infinity). 
The neutrino emission and annihilation is concentrated near the black hole, 
where accretion is expected to be quasi-steady.
$\dE$ depends on four parameters that 
specify the steady disk model: $\dM$, $\alpha$, $M$, and $a$. 

\medskip

The energy deposition rate $\dE$ was estimated in \cite{Popham}, 
approximating geodesics by straight lines. 
Fully relativistic calculations were made for several toy models, 
in particular for disks or tori with uniform temperature or other 
arbitrary distribution of temperature or entropy (see \cite{Birkl} and 
refs. therein).
Recently, the relativistic calculation for a realistic disk around a 
spinning black hole has been done (Zalamea \& Beloborodov, in preparation). 
$\dq$ has been obtained everywhere around the black hole, 
including its ergosphere, and the dependence of $\dE$ 
on $\dM$, $\alpha$, $M$, and $a$ has been determined.

\medskip

Besides tracing the geodesics, this calculation involves a model for 
neutrino and anti-neutrino spectra emitted by the disk.
Using the results of \cite{CB}, it is straightforward to evaluate the 
spectra from the transparent zone of the disk. 
It is more difficult to find the spectrum that emerges from the 
opaque zone, because the neutrino transport in this zone depends on the 
unknown vertical distribution of viscous heating. Various assumptions 
may be made about this distribution 
\cite{diMatt,Sawyer,Ramirez,Rossi2,Kawabata},
including a strong heating of the magnetic corona above the disk. 
Note that the corona of a hyper-accretion disk is always in thermodynamic 
equilibrium, and its temperature $T_c$ is determined by its thermal energy 
density $U_c=U_\gamma+U_e\approx 3a_rT_c^4$ where 
$a_r=7.56\times 10^{-15}$~erg~cm$^{-3}$~K$^{-4}$ is the radiation constant. 
$U_c$ is generally smaller than the energy density inside the disk.
Therefore, relocating the heating from the disk to its corona
cannot significantly increase the energies of emitted neutrinos.

\medskip

Fortunately, a robust estimate can be obtained for $\dE$ in spite of 
the uncertainty in the vertical structure of the disk.
It is easy to see that all detailed models of neutrino spectrum formation 
must predict practically the same rate of $\nu\bar\nu$ annihilation
above a {\it neutrino-cooled} disk. For such a disk,
neutrinos carry away a fixed energy flux $F^-\approx F^+$
where $F^+\sim 3\dM\OmK^2 S(r)/8\pi$ is the rate of viscous heating.
Therefore, transfer models that predict a higher average energy of emitted 
neutrinos, $\Eav$, must also predict a lower number density of the neutrinos 
above the disk, $n\sim F^-/\Eav c\propto \Eav^{-1}$.
Since the annihilation cross section $\sigma_{\nu\bar\nu}\propto \Eav^2$, 
one finds that the reaction rate
$\dot{n}_{\nu\bar\nu}\sim c \sigma_{\nu\bar\nu} n_\nu n_{\bar\nu}$ 
is independent of $\Eav$. The energy deposition rate
$\dq\sim\dot{n}_{\nu\bar\nu}\Eav$ is proportional to $\Eav$.
It cannot be changed without a substantial change in temperature
(or electron chemical potential) of the neutrino source, which 
would require a huge change in energy density and therefore is hardly possible.

\medskip

Detailed calculations confirm that $\dE$ weakly depends on the details of 
the vertical structure and $\nu,\bar{\nu}$ transfer in the disk.
Consider two extreme models for the opaque zone.
{\bf Model~A:} Neutrinos $\nu_e$ and $\bar\nu_e$ are emitted with 
the same distributions as found inside the disk (same temperature $T$ and
chemical potential $\mu_\nu$). 
The distribution normalization is, however, reduced compared with the 
thermal level inside the disk, so that the emerging emission carries away 
the known energy fluxes $F_{\nu}$ and $F_{\bar\nu}$ that are found in \cite{CB}.
{\bf Model~B:} Neutrinos are emitted with a thermal Fermi-Dirac spectrum 
with chemical potential $\mu_\nu=0$ and the effective surface temperature 
$\Teff$. The effective temperature is defined by $(7/8)\sigma\Teff^4=F^-$,
where $F^-=F_\nu+F_{\bar\nu}$ is the energy flux from one side of the disk 
and $\sigma=a_rc/4$ is Stefan-Boltzmann constant. The factor $7/8$ is 
determined by the difference between Plank and Fermi-Dirac distributions 
and the fact that the disk emits only $\nu_e$ and $\bar{\nu}_e$ --- 
the emission of other neutrino species is negligible.

\begin{figure}
\includegraphics[height=0.43\textheight]{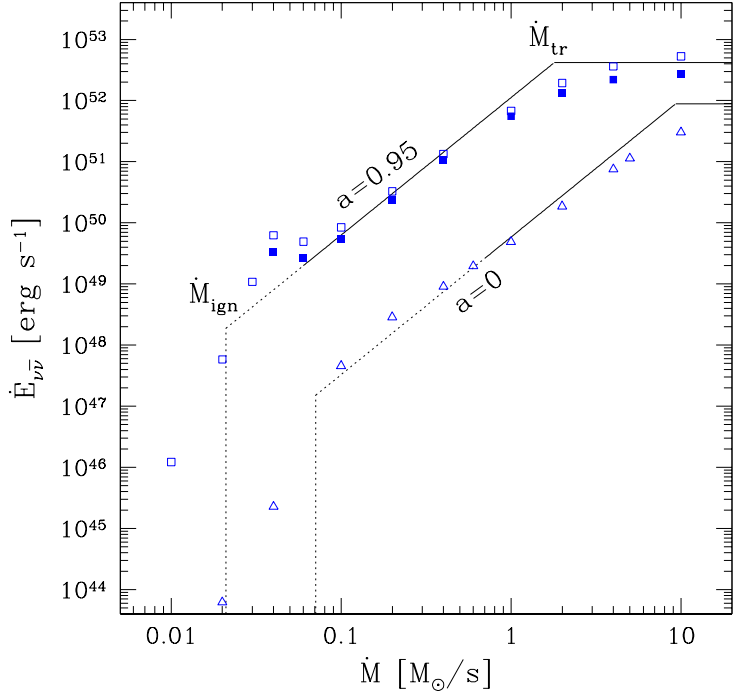}
\caption{Total energy deposition rate due to $\nu\bar\nu$ annihilation
outside the black-hole horizon, $\dot{E}_{\nu\bar\nu}$, as a function 
of the disk accretion rate, $\dM$. The two characteristic accretion rates
$\dMign$ and $\dMtr$ depend on the viscosity parameter $\alpha=0.1$
(see eq.~\ref{eq:pw}); $\alpha=0.1$ is chosen in this figure.
The black hole is assumed to have mass $M=3$\,M$_\odot$.
$\dE$ strongly depends on the spin parameter of the black hole; 
the numerical results are shown for two cases:
$a=0$ (triangles) and $a=0.95$ (squares). The uncertainty in the vertical 
structure of the accretion disk leads to a small uncertainty in $\dE$
as illustrated by two extreme models: Model~A (open symbols) and Model~B
(filled symbols), see the text for details. The results of both models
are well approximated by simple Model~C that is given by 
Eq.~(\ref{eq:Edot}) and also shown in the figure, by solid line for
$\nu$-opaque disks ($\dM>\dMop$) and by dashed line for $\nu$-transparent
disks ($\dM<\dMop$). (From Zalamea \& Beloborodov, in preparation.)}
\end{figure}

The temperature $T$ inside an opaque disk exceeds $\Teff$ by factor
$T/\Teff\sim \tau_\nu^{1/4}$ when the neutrino optical depth $\tau_\nu\gg 1$.
The neutrino chemical potential is modest and, approximately, $\Eav\propto T$.
Hence, even the extreme Model~A gives a moderate enhancement of $\dE$, 
by the factor $T/\Teff$. Zalamea \& Beloborodov (in preparation)
calculated $\dE$ in both Models A and B. The results of numerical 
calculations are shown in Fig.~6 and demonstrate that the difference 
between the two models is indeed small. 

\medskip

It is instructive then to consider {\bf Model C}: same as Model~B except 
that $F^-=F^+$ is assumed at all radii.
The assumption is clearly incorrect outside the region $\rtr<r<\rign$. 
Nevertheless, this simplest model gives a good approximation to $\dE$
in a broad range $\dMign<\dM<\dMtr$ (Fig.~6).\footnote{
    The deviation of Model~B from Model~C at $\dM\sim 2\dMign$ is 
    caused by the overshooting of $F^-$ above $F^+$, which happens just 
    inside of $\rign$ (see Fig.~7 and 16 in \cite{CB}). As hot matter 
    accretes into the neutrino-cooled region, its stored heat is quickly 
    emitted with $F^-/F^+$ reaching $\sim 2$ at $r\approx \rign/2$. For 
    disks with $\rign/2\sim$ a few $r_g$, this leads to the enhancement of 
    $\dE$ by the factor $\sim 2^{9/4}$ compared with Model~C that assumes
    $F^-=F^+$.}
Note that $\dE$ in Model~C is explicitly independent of $\alpha$. 
However, the range of $\dM$ where Model~C is applicable depends on $\alpha$ 
(Eq.~\ref{eq:pw}).  

\medskip

The scaling of $\dE$ with $\dM$ is easy to evaluate analytically.
The effective surface temperature is related to $F^-\approx F^+$ by 
$\Teff^4\propto F^-\propto\dM$. The neutrino number density above the disk 
is proportional to $\Teff^3\propto\dM^{3/4}$. The annihilation cross-section 
$\sigma_{\nu\bar\nu}\propto \Teff^2$ (assuming $k\Teff>m_ec^2$). 
Hence the energy deposition rate $\dE\propto\Teff^9\propto \dM^{9/4}$, and 
one can write
\be
\label{eq:Edot}
   \dE=\dot{E}_0(a)\left(\frac{\dM}{{\rm M}_\odot{\rm ~s}^{-1}}\right)^{9/4},
  \qquad \dMign<\dM<\dMtr.
\ee
The normalization factor $\dot{E}_0$ depends on the black hole spin $a$ and 
must be calculated numerically. For example, $a=0.95$ gives 
$\dot{E}_0\approx 10^{52}$~erg~s$^{-1}$, which implies the efficiency
$\epsilon=\dE/L\approx 0.05(\dM/{\rm M}_\odot{\rm ~s}^{-1})^{5/4}$.
It is much larger than the corresponding 
value for a non-rotating black hole, by two orders of magnitude. 

\medskip

The strong dependence of $\dE$ on $a$ may be seen from the following rough 
estimate. The neutrino luminosity $L$ peaks at $\rp$ that is a few times 
the inner radius of the disk, $\rms(a)$ --- the marginally stable orbit, 
which is determined by $a$. 
The luminosity depends on $\rms$ approximately as $\rms^{-1}$, and 
$\Teff$ at $\rp$ scales as $(L/\rp^2)^{1/4}\propto \rms^{-3/4}$. The energy
deposition rate $\dq$ scales as $\Teff^{9}$ and 
$\dE$ scales as $\rms^3\dq$, which yields $\dE\propto \rms^{-15/4}$. 
Then the reduction in $\rms$ by a factor of 3 (as $a$ increases from 0 to 
0.95) gives a factor of 60 in $\dE$. This estimate neglects the fact that 
the gravitational bending of neutrino trajectories is stronger for smaller 
$\rms$. Stronger bending implies a larger average angle between neutrinos
and leads to an additional enhancement of $\dE$. Therefore, a steeper 
dependence of $\dE$ on $\rms$ is found in numerical simulations.
A simple power law $\dE\propto\rms^{-4.7}$ is an excellent approximation 
to the numerical results for $0<a<0.95$ which corresponds to 
$r_g<\rms<3r_g$ (Zalamea \& Beloborodov, in preparation).

\medskip

Note that $\dE$ is defined as the {\it total} energy deposition rate 
outside the event horizon (including the ergosphere). A significant 
fraction of the created $e^\pm$ plasma must fall into the black hole, 
and only the remaining fraction of $\dE$ will add energy to the jet. 
This fraction depends on the plasma dynamics outside the disk, which is 
affected by magnetic fields and is hard to calculate without additional 
assumptions.


\section{Low-angular-momentum disks in collapsars}

The quasi-spherical accretion flows in collapsars create a centrifugally 
supported disk if the circularization radius of the flow is sufficiently
large, $r_{\rm circ}\sim 10r_g$.
A smaller disk may not be centrifugally supported and then will accrete
on a free-fall timescale \cite{BI}. It accretes so fast (super-sonically) 
that the effects of viscosity can be neglected. A steady model of this 
``mini-disk'' was constructed in \cite{BI} and 2D time-dependent 
hydrodynamical simulations were performed in \cite{Lee2}. 

\medskip

The mini-disk can be thought of as a caustic in the equatorial plane of 
a rotating accretion flow. It absorbs the feeding infall, and this 
interaction releases energy, making the accretion radiatively efficient.
With increasing angular momentum, the size of the disk grows up to $14r_gc$,
at which point the centrifugal barrier stops accretion, so that it
can proceed only on a viscous timescale. Thus, the mini-disk model fills 
the gap between two classical regimes of accretion
--- spherical ($l<r_gc$) and standard accretion disk ($l\gg r_gc$)
--- and is qualitatively different from both. 

\medskip

The calculations of \cite{BI} were limited to the case of a Schwarzschild 
black hole. Recently, the model has been extended to the case of 
a Kerr black hole (Zalamea \& Beloborodov, in preparation). 
Fig.~7 shows the range of angular momenta that lead to mini-disk
formation around a black hole of spin $0<a<1$. The critical angular momentum
for viscous disk formation sets the maximum radius of a mini-disk. This 
radius is $\approx 14r_g=28GM/c^2$ for $a=0$ and $\approx 5r_g$ for $a=0.95$.

\begin{figure}
\includegraphics[height=0.35\textheight]{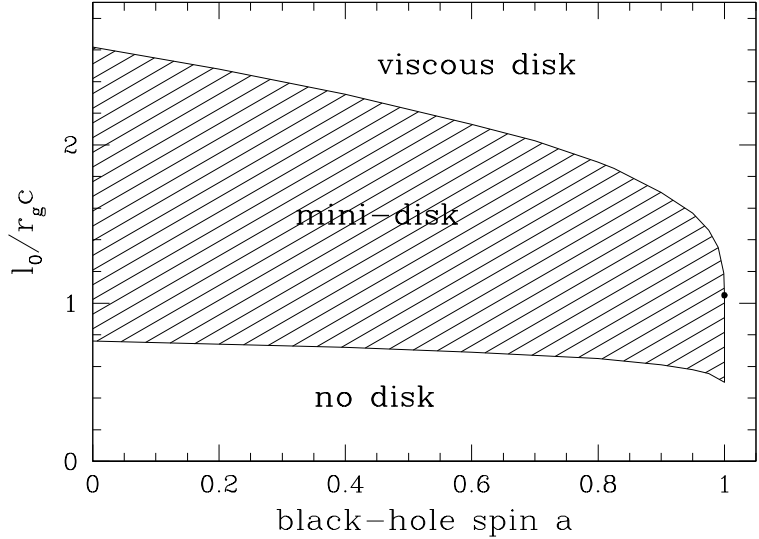}
\caption{The shaded region shows the range of angular 
momenta $l_0$ that form a mini-disk with insufficient centrifugal support, 
leading to accretion on the free-fall timescale. $l_0$ is defined as the 
angular momentum of the accretion flow in the equatorial plane;
angular momentum decreases toward the polar axis (shells $r=const$ are
assumed to have a uniform angular velocity $\Omega\ll\OmK$ at $r\gg r_g$).
(From Zalamea \& Beloborodov, in preparation).  }
\end{figure}

\medskip

The mini-disk is sandwiched by shocks through which the 
infalling matter enters the disk. The model of \cite{BI} assumes that 
the cooling timescale of the postshock material is sufficiently short,
shorter than the accretion timescale in the disk.
Let us check if this assumption can be valid for collapsars.
The postshock mass density and energy density can be estimated as 
\be
  \rho\sim 10^9\xi\dMM M_3^{-2}\left(\frac{r}{r_g}\right)^{-3/2}
     {\rm ~g~cm}^{-3}, \qquad 
   U\sim 0.2 \rho c^2\left(\frac{r}{r_g}\right)^{-1}.
\ee
Here $\xi\sim 2$ is the compression in the shock (note that the shock 
dissipates only the normal component of the infall velocity); 
$\dM_{0.1}\equiv \dM/0.1\,{\rm M}_\odot{\rm ~s}^{-1}$ and $M_3\equiv M/3\,{\rm M}_\odot$. 
The postshock matter has $\mu_e/kT<1$, and $U$ is dominated 
by radiation and $e^\pm$ pairs, which implies $U\approx 3a_rT^4$.
The postshock temperature is then
$T\approx 5.3\times 10^{10}\,\rho_9^{1/4}(r/r_g)^{-1/4}$~K. 
Disintegration of nuclei in the shock consumes only $\sim 10^{-2}$ of the 
energy released at $r\sim r_g$, so the postshock matter can be cooled only 
by neutrino emission. Neutrino emission is dominated by two processes:
(i) capture reactions~(\ref{eq:reactions}) provide cooling rate 
$\dot{q}_c\approx 9\times 10^{32}T_{11}^6\rho_9$~erg~cm$^{-3}$~s$^{-1}$, and 
(ii) $e^\pm$ annihilation $e^++e^-\rightarrow \nu+\bar{\nu}$ provides 
$\dot{q}_\pm\approx 3.6\times 10^{33}T_{11}^9$~erg~cm$^{-3}$~s$^{-1}$ 
(see e.g. \cite{Yakovlev}). This gives
\be
 \dot{q}_c  \approx 2\times 10^{31}\rho_9^{5/2}
    \left(\frac{r}{r_g}\right)^{-3/2} \;\frac{\rm erg}{{\rm cm}^3{\rm ~s}},
\qquad
 \dot{q}_\pm\approx 10^{31}\rho_9^{9/4}
    \left(\frac{r}{r_g}\right)^{-9/4} \;\frac{\rm erg}{{\rm cm}^3{\rm ~s}},
\ee
Approximating the total $\dot{q}=\dot{q}_c+\dot{q}_\pm\sim \dot{q}_c$,
one finds
\be
 \frac{\dot{q}\;\tacc}{U}\sim 10^{-2}\rho_9^{3/2}\left(\frac{r}{r_g}\right)M_3
   \sim 10^{-2}\xi^{3/2}\dMM^{3/2}\left(\frac{r}{r_g}\right)^{-5/4}M_3^{-2},
\ee
where $\tacc\sim 10^{-4}(r/r_g)^{-3/2}M_3$~s. The mini-disk is 
neutrino-cooled if $\dot{q}\;\tacc/U\simgt 1$, which requires a high 
accretion rate, comparable to M$_\odot$~s$^{-1}$. Thus, only high-$\dM$ 
mini-disks are sandwiched by radiative shocks that stay near the equatorial 
plane in the innermost region of the accretion flow. A large neutrino 
luminosity, up to $\sim 0.1\dM c^2$, is produced by such disks. 

\medskip
 
For smaller accretion rates, the postshock matter is unable to cool on the 
free-fall timescale, and the neutrino luminosity from the inner region is 
suppressed by the factor $\dot{q}\tacc/U < 1$.
Then a hot low-angular-momentum bubble must grow around the black hole. 
Mass flows into the bubble through the shock front that expands to 
$r\gg r_g$. Such a bubble is observed in low-$\dM$ simulations in \cite{Lee2}.
It resembles the bubble around viscous disks in the models of 
\cite{MacFadyen,Igumen}, except for a slower rotation, less centrifugal 
support, and faster accretion. The shock expansion can be stopped when it 
approaches $\sim 40r_g=4\times 10^7M_3$~cm. Up to this radius, the 
postshock temperature, $T\approx 7\times 10^9\dMM^{1/4}(r/40r_g)^{-5/8}$~K, 
is high enough to disintegrate nuclei at the density
$\rho\sim 4\times 10^6\dMM (r/40r_g)^{-3/2}$~g~cm$^{-3}$.
As the shock expands to $40r_g$, its energy decreases to
$GMm_p/r\approx 12(r/40r_g)^{-1}$~MeV per nucleon, and a large fraction of 
this energy is consumed by disintegration (8~MeV per nucleon); 
therefore, the shock stalls. 


\section{Spreading of viscous disks and nuclear burning}

Formation of a viscous disk with $r_{\rm circ}>10 r_g$ implies that most of 
the angular momentum of accreting matter will be stored outside the black 
hole, in a viscously spreading ring. When matter supply to the disk stops, 
accretion will proceed from this ring. 
At any time $t$, the characteristic size of the ring $R(t)$ is where its 
mass peaks. Alternatively, $R$ can be defined by $\Jd=(GMR)^{1/2}\Md$, 
where $\Jd$ is the angular momentum carried by the disk and $m$ is its mass.

\subsection{Spreading of merger disks}

Immediately after the merger, the characteristic size of the debris disk 
is $R_0\sim 10^7$~cm, and its initial mass $m_0$ may be as large as 
$\sim 0.1\,$M$_\odot$.\footnote{
    The mass of the debris disk is sensitive to the parameters of the binary 
    system before the merger, in particular to the mass ratio and the spins 
    of the two companions, see e.g. \cite{Lee1} for a review.} 
Its viscous evolution starts on a timescale 
$t_0=(\alpha\OmK)^{-1}(H/r)^{-2}\simlt 0.1(\alpha/0.1)^{-1}$~s, with
accretion rate $\dM_0\sim m_0/t_0$ that can exceed $1$\,M$_\odot$~s$^{-1}$.
The disk is initially hot and $\nu$-opaque (cf. Fig.~2); 
its nuclear matter is composed of free nucleons $n$ and $p$. 

\medskip

The initial accretion phase lasts $\sim t_0$. Following this stage, the disk 
mass $m(t)$ is reduced and its radius $R(t)$ grows to conserve the angular 
momentum, $J=m(GMR)^{1/2}\approx const$, which implies $m\propto R^{-1/2}$. 
Several important changes occur in the disk as it spreads to 
$R\sim 10^2r_g\approx 10^8$~cm:

\begin{list}{$\circ$}{\setlength{\leftmargin}{5mm}} 

\item Temperature $T$ and electron chemical potential $\mu_e$ in the outer 
region $r\sim R$ decrease to $\sim 1$~MeV. As a result, the disk material at 
$r\sim R$ is not $\nu$-cooled anymore: $R(t)$ exits the neutrino-cooled 
region $r<\rign$ on the $\dM-r$ diagram (Fig.~2). The viscously produced 
heat outside $\rign$ is stored and advected by the spreading accretion disk. 
The spreading matter is then marginally bound to the black hole, 
$c_s\sim \vK$.

\item Electrons become non-degenerated. Pressure is not dominated
by neutrons anymore: it is dominated by radiation and $e^\pm$ pairs,
$P\approx P_\gamma+P_\pm\approx a_rT^4$. 

\item $Y_e$ freezes.

\item Nuclear burning occurs: free nucleons $n$ and $p$ recombine into 
$\alpha$ particles. This process releases energy of 7~MeV per nucleon,
comparable to the binding energy $GMm_p/r$, and unbinds most of the disk 
matter, ejecting it in a freely expanding wind.\footnote{
    In addition to nuclear burning and viscous heating, the matter is heated 
    by neutrinos emitted at $r\sim r_g$ (the mass accretion rate by the black 
    hole is still significant when $R(t)$ approaches $10^8$~cm). The energy 
    deposited by neutrinos in the advective zone of the spreading disk, 
    $r\simgt 50r_g$ is comparable to the viscously dissipated energy in 
    this zone.}  

\end{list}

\noindent All these changes happen as $R(t)$ grows from $\sim 50$ to $\sim 100r_g$. 

\medskip 

A one-zone model of the spreading disk is calculated in a recent work 
\cite{Metzger1}. Let us estimate here one characteristic radius 
$\rfr$ at which $\tvisc=\tw$. Here $\tw$ is the time of conversion
$n\leftrightarrow p$ through reactions~(\ref{eq:reactions}) \cite{Imsh}.
At~$\rfr$, pressure is already becoming dominated by radiation and 
non-degenerate $e^\pm$ pairs. On the other hand, $n$ and $p$ have not yet 
recombined. The timescales $\tw$ and $\tvisc$ are then given by 
\be
  \tw\approx 70\left(\frac{kT}{m_ec^2}\right)^{-5}\;{\rm~s}, \qquad
   \tvisc\approx \frac{1}{\alpha\OmK(r)}\,\left(\frac{H}{r}\right)^{-2}.
\ee
Using the hydrostatic balance $P/\rho=(H/r)^2\vK^2$ with 
$\rho\approx m/4\pi r^2H$, one finds 
\be
   kT\approx 0.60\; h^{1/4}M_3^{1/4}\,\frac{m_{32}^{1/4}}{R_8}{\rm ~~MeV},
\qquad
  \frac{\mu_e}{kT}\approx 1.0\;Y_e\,h^{-7/4}M_3^{-3/4}m_{32}^{1/4},
\ee
where $h\equiv 2H/R$, $M_3\equiv M/3$\,M$_\odot$, 
$m_{32}\equiv m/10^{32}{\rm ~g}=(m/0.05\,$M$_\odot)$, and $R_8\equiv R/10^8$cm.
(Note that $\mu_e/kT$ depends on $m$ and $h$ only, not $R$.) This gives
\be
  \frac{\tw}{\tvisc}\approx 17\; h^{3/4}\alpha_{0.1}M_3^{-3/4} R_8^{7/2}
   m_{32}^{-5/4}
   \;\;\;  
\ee
\be 
\label{eq:rfr}
  \Rightarrow \;\;
  \rfr\approx 4.5\times 10^7 h^{-3/14}\alpha_{0.1}^{-2/7}M_3^{3/14}
              m_{32}^{5/14}{\rm ~~cm},
\ee
where $\alpha_{0.1}=\alpha/0.1$.  The density and temperature of the disk 
at $r=R=\rfr$ are
\be
   kT_\star\approx 1.3\, h^{13/28}\alpha_{0.1}^{2/7}M_3^{1/28}
                   m_{32}^{-3/28} {\rm ~MeV},
\ee
\be
  \rho_\star\approx 2\times 10^8\,h^{-5/14}\alpha_{0.1}^{6/7} M_3^{-9/14} 
                     m_{32}^{-1/14} {\rm ~g~cm}^{-3}.
\ee
At this temperature and density matter is close to the neutronization line
$kT_n(\rho)=1.04\;(\rho/10^8{\rm ~g~cm}^{-3})^{1/2}$~MeV (Fig.~2), 
\be
  \frac{T_\star}{T_n(\rho_\star)}
  \approx 0.9\, h^{9/14}\alpha_{0.1}^{-1/7}M_3^{5/14}m_{32}^{-1/14}. 
\ee
Hence the {\it equilibrium} value of $Y_e$ at $\rfr$ is $Y_e^\star\approx 0.5$.
The actual $Y_e$ in the spreading disk gradually freezes out as $R(t)$ passes 
through $\rfr$ and its asymptotic value after the transition can differ from 
$Y_e^\star$. The freeze-out $Y_e\sim 0.3$ is found in \cite{Metzger1} 
using a dynamical model for the spreading disk.
Note that the model describes the {\it average} value of $Y_e$.
Viscous spreading is a random diffusion process, so different elements of 
the disk spend different times near $\rfr$, and a longer residence time 
at $\rfr$ gives a higher $Y_e$. One can therefore expect a mixture of 
different $Y_e$ in the spreading disk, with a dispersion 
$\Delta Y_e/Y_e\sim 1$ around the average value. 

\medskip

Soon after passing $\rfr$ this mixture is heated by nuclear recombination 
and ejected in a wind of duration $\sim\tvisc(R=10^8{\rm ~cm})$.
Subsequent nucleosynthesis in the expanding ejecta produces diverse
radio-active elements, including some with a long life-time. 
Their decay can make the ejecta visible to a distant observer \cite{Li}.
In particular, material with $Y_e\approx 0.5$ will synthesize $^{56}$Ni.
$^{56}$Ni decays when the ejecta expand so much that their thermal radiation 
can diffuse out and escape to observer, producing an optical flash similar
to normal supernovae.

\subsection{Spreading of collapsar disks}

The collapsar disks are continually fed by the
infalling stellar matter during a long time $\tinf\sim 10$~s (and longer, 
with a decreasing infall rate). The model posits that the angular momentum 
of the infall, $\linf$, is sufficiently large to form a viscous disk 
\cite{Woosley,MacFadyen}, e.g., the circularization radius of the infall in 
the numerical model of \cite{MacFadyen} is $\rc\simlt 30r_g$. The accretion 
timescale at this radius, $\tvisc\sim 3\times 10^{-2}\alpha_{0.1}^{-1}$s 
is much shorter than $\tinf$.
This led \cite{MacFadyen} and many subsequent works to picture a low-mass 
disk, $\Md\sim \tvisc\dM\sim 3\times 10^{-3}\dM$ that is continually 
drained into the black hole and re-filled with fresh infalling matter.

\medskip

The picture of a low-mass viscous disk is, however, implausible.
Conservation of angular momentum requires the following: 
(1) The disk spreads during $\tinf$ to a radius 
$R\sim 3\times 10^2r_g(\alpha/0.1)^{-1}$
where the viscous timescale is comparable to $\tinf$.
(2) The disk accumulates mass $m$ that
carries angular momentum $J=\Jtot-\Jacc$. Here $\Jtot=\Macc\linf$ is 
the total angular momentum processed by the collapsar disk,
$\Jacc\sim\Macc r_gc$ is the angular momentum accreted by the black hole,
and $\Macc\sim\dM\tinf\sim $M$_\odot$ is the mass accreted through the disk. 
Since $\Jtot>\Jacc$ for any viscous disk and usually $\Jtot\gg\Jacc$,
such disks must store $\Jd\sim \Jtot$ (unless almost all angular momentum 
is carried away by a wind). This implies that the disk accumulates
the mass,
\be
   m\approx\Macc\frac{\lK(R)}{\linf}=\Macc\left(\frac{R}{\rc}\right)^{1/2}
    \sim {\rm M}_\odot.
\ee
The disk mass may be much smaller than this estimate only if $\linf$ is 
so small that $\Jtot\approx\Jacc$, which leaves $J\ll\Jtot$ for the disk. 
This condition leads, however, to the inviscid mini-disk described in 
the previous section. A low-mass {\it viscous} disk could form only if 
$\linf$ is fine-tuned toward the boundary between the viscous and mini-disk 
accretion regimes (cf. Fig.~7).\footnote{
    The very small $m\sim 0.003$\,M$_\odot$ found in the simulations of 
    \cite{MacFadyen} may be the result of the imposed absorbing boundary 
    condition at $r_{\rm in}=50{\rm ~km}\approx 5r_g$, which is $\sim 5$ 
    times larger than the true inner radius of the disk, $\rms\sim r_g$. 
    The large $r_{\rm in}$ implies an artificially large $\Jacc$, which 
  happens to be nearly equal to $\Jtot$ in the model, permitting $J\ll\Jtot$. 
  In addition, a (small) fraction of $\Jtot$ is carried away by the wind. }

\medskip

Spreading of viscous disks in collapsars through a radius $\sim 10^2r_g$ is 
accompanied by significant changes, similar to the evolution of merger disks 
described above. In particular, matter acquires a positive Bernoulli constant 
as a result of viscous, nuclear, and neutrino heating. The infalling material 
of the progenitor star exerts an external ram pressure on the disk, and
can confine the disk initially, but eventually the disk pressure must win 
and its matter will expand with a velocity $\sim 2\times 10^9$~cm/s. 
The ejected mass $\sim 10^{33}$~g carries the energy $E\sim 10^{52}$~erg 
and will explode the outer parts of the star. To a first approximation, 
the expansion of the outer disk may be described as a thermal explosion 
driven mainly by nuclear burning of $n$ and $p$ into $\alpha$ particles. 
A fraction of the unbound disk matter will turn into $^{56}$Ni and should
create a bright supernovae-like event.

\subsection{Self-similar spreading at late stages}

The disk matter that has spread beyond $\sim 10^8$~cm is largely unbound 
and ejected, however, some matter remains bound and rotating in a remnant
disk. Its mass is hard to estimate; it could be as large as 
$\sim 0.1$\,M$_\odot$ for collapsars and $\sim 0.01$\,M$_\odot$ for mergers.
This remnant is composed of recombined nucleons, and further fusion 
reactions are not a significant source of energy (compared with 
the virial/gravitational energy). The central source of neutrinos switches 
off as $\dM$ drops, so neutrino heating is also insignificant. 
This advective remnant disk will continue to spread viscously to larger
radii, gradually draining its mass into the black hole and possibly 
losing mass to a wind. 

\medskip

If the mass loss through a wind is small, the spreading enters a simple
self-similar regime such that $\Jd=const$, $R(t)$ grows as a power-law 
with time, while $\dM(t)$ and $m(t)$ decrease as power-laws with time.
Detailed self-similar models of this type were studied, see e.g. 
\cite{Lynden,Ogilvie}. The advective disk has a scale-height $H\sim r$, 
sound speed $c_s\sim \vK=(GM/r)^{1/2}$, and $\Omega\sim \OmK$. Its kinematic 
viscosity coefficient is $\nu\sim \alpha c_s H\sim\alpha \vK r=\alpha \lK$, 
where $\lK(r)=(GMr)^{1/2}$. The disk spreading is a diffusion process 
described by
\be 
  R^2(t)\sim \nu(R) t\sim \alpha(GMR)^{1/2}t  \qquad \Rightarrow \quad\;
   R(t)\sim \alpha^{2/3}(GM)^{1/3} t^{2/3}
       \sim R_0\left(\frac{t}{t_0}\right)^{2/3},
\ee
where subscipt ``0'' refers to an initial reference moment of time $t_0$.
The disk mass $m(t)$ is then found from the condition 
$\Jd=(GMR)^{1/2}m=const$,
\be
   m(t)=\frac{\Jd}{(GMR)^{1/2}}
      \sim m_0\left(\frac{R}{R_0}\right)^{-1/2}
      \sim m_0\left(\frac{t}{t_0}\right)^{-1/3},
\ee
and the accretion rate is given by
\be
   \dM(t)\sim \frac{m}{t}\sim \dM_0\left(\frac{t}{t_0}\right)^{-4/3}.
\ee

\medskip

This self-similar solution may not apply if the disk loses mass through a 
wind and \cite{Metzger1} consider solutions that include the wind.
In general, advective disks are only marginally bound by the gravitational
field of the black hole and their Bernoulli constant can be positive
\cite{Narayan2,Blandford}. This is expected to cause a strong wind.
On the other hand, bound solutions with a negative Bernoulli constant were
found for spreading advective disks \cite{Ogilvie}. The mass loss through 
a wind then depends on the poorly understood vertical distribution of viscous 
heating inside the disk and the behavior of the magnetic field above the disk.


\section{Conclusions}

Hyper-accretion disks are formed from matter with a modest angular momentum,
with circularization radius $r_{\rm circ}$ well inside 
$10^2 r_g\approx 10^8$~cm. Matter with $r_{\rm circ}\simgt 10r_g$ is
supported by the centrifugal barrier, and its accretion is driven by 
viscous stresses on a timescale $t_0\sim 0.1(\alpha/0.1)^{-1}$~s, where 
$\alpha\sim 0.01-0.1$ is the viscosity parameter.
These viscous disks are dense and hot, and emit copious 
neutrinos as long as the accretion rate $\dM$ exceeds 
$\dMign\sim 0.03(\alpha/0.1)^{5/3}\,$M$_\odot$~s$^{-1}$.

\medskip

Neutrino annihilation above the disk deposits a significant energy that can 
power GRB explosions. In contrast to previous expectations, the rate of 
energy deposition $\dE$ is found to be not sensitive to the details 
of neutrino transport and the vertical structure of the accretion disk.
It is given by $\dE\approx \dot{E}_0(\dM/{\rm M}_\odot{\rm ~s}^{-1})^{9/4}$
in a broad range of accretion rates $\dMign\simlt \dM \simlt \dMtr$
(eq.~\ref{eq:pw}). The normalization factor $\dot{E}_0$ is very sensitive to 
the black hole spin $a$. For instance, $\dot{E}_0\approx 10^{52}$~erg~s$^{-1}$ 
is found for a black hole with $a=0.95$, which is two orders of magnitude 
larger than $\dot{E}_0$ for the Schwarzschild case $a=0$.

\medskip

Remarkably, all neutrino-cooled viscous disks regulate themselves to a 
characteristic state such that electrons are mildly degenerate, 
$Y_e\sim 0.1$, and free neutrons dominate the pressure in the disk \cite{CB}. 
The neutron-rich matter may contaminate the jet from the accreting black hole
and get ejected with a high Lorentz factor. Then the gradual decay of 
the ejected neutrons affects the global picture of GRB explosion on scales 
up to $10^{17}$~cm, where the GRB blast wave is observed 
\cite{Derishev,B03b,Rossi3}.  

\medskip

The disk size $R$ grows with time as a result of viscous spreading.
Most of the disk mass $m(t)$ resides near $R(t)$ and its state is not 
described by the steady model (which remains valid at radii $r<R$).
Instead, it is described by a markedly different spreading solution.
In particular, as the disk spreads to $\sim 100r_g$ it is heated both 
viscously and by the nuclear burning of free nucleons into helium.\footnote{
    In contrast, when matter {\it accretes} through $\sim 10^2r_g$
    (as in the steady-state model), this process is reversed: 
    helium is disintegrated, which leads to {\it cooling}.} 
As a result, the disk is disrupted before it spreads much beyond $10^8$~cm: 
the heated flow acquires a positive Bernoulli constant and gets unbound.

\medskip

Then most of the disk mass $m$ is ejected with a velocity 
$\sim 0.1$~c and a total energy $\sim 10^{52}(m/{\rm M}_\odot)$~erg.
A fraction of the ejected matter acquires $Y_e\approx 0.5$, which favors 
the synthesis of $^{56}$Ni as the ejecta expand and their temperature drops. 
The ensuing gradual decay of $^{56}$Ni should produce a visible optical 
flash on a week timescale --- a supernova-like event. The flash is expected 
to be especially bright for collapsars that develop massive spreading 
accretion disks. A similar (but weaker) flash should be 
produced by the spreading disks around a merged binary; the ejected mass can 
be a few orders of magnitude smaller in this case.

\medskip

$^{56}$Ni-rich matter may also be ejected from the {\it inner}, 
geometrically thin, neutrino-cooled disk. This can occur 
if the inner disk produces a strong wind \cite{Pruet,Barzilay,Metzger2,Surman}.
Such winds are modeled as quasi-steady magnetized outflows, 
illuminated by neutrinos which can heat and de-neutronize the wind material.
The details of this plausible mechanism are uncertain because
the mass outflow rate and the asymptotic $Y_e$ in the wind is hard to 
predict with confidence --- both depend on the assumed 
MHD behavior of the disk and its corona. 

\medskip

Following the main burst, the accretion rate is determined by the amount 
of matter that remains bound and rotating around the black hole.\footnote{
    In collapsars, the fallback of the progenitor envelope is often assumed 
    to determine $\dM$ at late times. This assumption is invalid {\it if a 
    viscous disk has formed}. The disk stores and then ejects so much mass 
    and energy that it must explode the star. Even if the disk ejected 
    little mass/energy, the accretion of the envelope would be negligible 
    --- the relict disk would supply a larger accretion rate 
    $\dM\propto t^{-4/3}$. }
$\dM(t)$ decreases steeply when the disk spreads beyond $10^8$~cm and most 
of its matter is ejected, however, some matter remains bound and continues 
to accrete. The evolution of $\dM$ may be related to the observed puzzling 
features in the afterglow emission of GRBs. The afterglow is likely to be 
produced by the relativistic blast wave driven by the jet from the central 
engine. Its luminosity is determined by the energy and magnetization of the
jet as well as the density profile of the ambient medium at 
$r\sim 10^{15}-10^{17}$~cm. A long-lived jet of luminosity 
$L_{\rm jet}=\epsilon_{\rm jet}\dM c^2$
would certainly impact the afterglow emission.
However, current theories are unable to reliably predict the evolution
of $\dM$ and $\epsilon_{\rm jet}$. For instance, one could speculate
that the jet switches off abruptly as $\dM$ decreases below a threshold,
which causes the observed steep decay in the afterglow light curves.

\medskip

While the mechanism of the relativistic jet and its evolution with $\dM(t)$
remain uncertain, the non-relativistic massive ejecta with $v\sim 0.1c$
is a robust consequence of viscous-disk accretion. Viscous disks certainly
form in merger events. The standard collapsar model also assumes the 
formation of a viscous disk, but this case is less certain. The minimum 
angular momentum needed to form a disk is $\sim r_gc$, and collapsars were 
proposed as rare events of stellar collapse with $l>r_g c$. Hence, 
statistically, the accretion flows in collapsars are likely to have small 
$l$ and their disks can be smaller than $\sim 10r_g$. Such mini-disks
are not centrifugally supported and accrete faster than viscous disks.
In contrast to the viscous regime, this low-angular momentum accretion 
leaves no remnant disk in the end of the core collapse, involves no 
viscous spreading, and may not eject much mass. However, it still can 
produce a powerful relativistic jet via the Blandford-Znajek mechanism
and/or neutrino heating near the rotation axis.


\begin{theacknowledgments}
I thank S.~Blinnikov for pointing out the work by Imshennik, Nadezhin, \& 
Pinaev \cite{Imsh}. This research was supported by NASA {\it Swift} grant.
\end{theacknowledgments}



\bibliographystyle{aipproc}   


\begin{thebibliography}{99}

\bibitem{Shibata1}
M.~Shibata, and K.~Taniguchi, 
``Merger of Black Hole and Neutron Star in Neneral Relativity: 
Tidal Disruption, Torus mass, and Gravitational Waves,''
\emph{Phys.~Rev.~D} \textbf{77}, id. 084015 (2008).

\bibitem{Etienne}
Z.~B.~Etienne, J.~A.~Faber, Y.~T.~Liu, S.~L.~Shapiro, K.~Taniguchi,  
and T.~W.~Baumgarte, 
``Fully General Relativistic Simulations of Black Hole-Neutron Star Mergers,''
\emph{Phys. Rev. D} \textbf{77}, id. 084002 (2008).

\bibitem{Liu1}
Y.~T.~Liu, S.~L.~Shapiro, Z.~B.~Etienne, and K.~Taniguchi, 
``General Relativistic Simulations of Magnetized Binary Neutron Star Mergers,''
\emph{Phys.~Rev.~D} \textbf{78}, id. 02401 (2008).

\bibitem{Ruffert}
M.~Ruffert, and H.-Th.~Janka, 
``Coalescing Neutron Stars - A Step Towards Physical Models. 
III. Improved Numerics and Different Neutron Star Masses and Spins,''
\emph{A\&A} \textbf{380}, 544--577 (2001).

\bibitem{Rosswog}
S.~Rosswog, ``Mergers of Neutron Star-Black Hole Binaries with Small Mass 
Ratios: Nucleosynthesis, Gamma-Ray Bursts, and Electromagnetic Transients,''
\emph{ApJ} \textbf{634}, 1202--1213 (2005).

\bibitem{Setiawan}
S.~Setiawan, M.~Ruffert, and H.-Th.~Janka, 
``Three-Dimensional Simulations of Non-Stationary Accretion by Remnant Black 
Holes of Compact Object Mergers,''
\emph{A\&A} \textbf{458}, 553--567 (2006).

\bibitem{Shibata2}
M. Shibata, Y. Sekiguchi, and R. Takahashia, 
``Magnetohydrodynamics of Neutrino-Cooled Accretion Tori around 
a Rotating Black Hole in General Relativity,''
\emph{Prog. of Theor. Physics} \textbf{118}, 257--302 (2007).

\bibitem{Woosley}
S.~E.~Woosley, ``Gamma-ray Bursts from Stellar Mass Accretion Disks around 
Black Holes,''
\emph{ApJ} \textbf{405}, 273--277 (1993).

\bibitem{MacFadyen}
A.~I.~MacFadyen, and S.~E.~Woosley,
``Collapsars: Gamma-Ray Bursts and Explosions in ``Failed Supernovae''
\emph{ApJ} \textbf{524}, 262-289 (1999).

\bibitem{Barkov}
M. Barkov, and S. S. Komissarov,
``Stellar Explosions Powered by the Blandford-Znajek Mechanism,''
\emph{MNRAS} \textbf{385}, L28--L32 (2008).

\bibitem{Piran}
T.~Piran, ``The Physics of Gamma-Ray Bursts,''
\emph{Rev. Mod. Phys.} \textbf{76}, 
1143--1210 (2005).

\bibitem{Balbus}
S.~A.~Balbus, and J.~F.~Hawley, 
``Instability, Turbulence, and Enhanced Transport in Accretion Disks,''
\emph{Rev.~Mod.~Phys.} \textbf{70}, 1--53 (1998).

\bibitem{Popham}
R.~Popham, S.~E.~Woosley, and C.~Fryer,
``Hyperaccreting Black Holes and Gamma-Ray Bursts,''
\emph{ApJ} \textbf{518}, 356--374 (1999).

\bibitem{Narayan1}
R.~Narayan, T.~Piran, and P.~Kumar, 
``Accretion Models of Gamma-Ray Bursts,''
\emph{ApJ} \textbf{557}, 949--957 (2001).

\bibitem{diMatt}
T.~Di~Matteo, R.~Perna, and R.~Narayan,
``Neutrino Trapping and Accretion Models for Gamma-Ray Bursts,''
\emph{ApJ} \textbf{579}, 706--715 (2002).

\bibitem{Kohri}
K.~Kohri, R.~Narayan, and T.~Piran, 
``Neutrino-dominated Accretion and Supernovae,''
\emph{ApJ} \textbf{629}, 341--361 (2005).

\bibitem{CB}
W.-X.~Chen, and A.~M.~Beloborodov, 
``Neutrino-cooled Accretion Disks around Spinning Black Holes,''
\emph{ApJ} \textbf{657}, 383--399 (2007).

\bibitem{Kawanaka}
N.~Kawanaka, and S.~Mineshige,
``Neutrino-cooled Accretion Disk and Its Stability,''
\emph{ApJ} \textbf{662}, 1156--1166 (2007).

\bibitem{Janiuk}
A.~Janiuk, Y.~Yuan, R.~Perna, and T.~Di~Matteo,
``Instabilities in the Time-Dependent Neutrino Disk in Gamma-Ray Bursts,''
\emph{ApJ} \textbf{664}, 1011-1025 (2007).

\bibitem{Rossi1}
E.~Rossi, P.~J.~Armitage, and K.~Menou,
``Microphysical Dissipation, Turbulence and Magnetic Fields in 
Hyper-Accreting Discs,''
\emph{MNRAS}, submitted (arXiv:0807.3547)

\bibitem{B99}
A.~M.~Beloborodov, ``Accretion Disk Models,''
in \emph{High Energy Processes in Accreting Black Holes}, 
edited by J.~Poutanen and R.~Svensson, ASP Conference Series 161, 
 1999, pp.~295--314

\bibitem{B98}
A.~M.~Beloborodov, ``Super-Eddington Accretion Discs around Kerr Black Holes,''
\emph{MNRAS} \textbf{297}, 739--746 (1998).

\bibitem{Imsh}
V.~S.~Imshennik, D.~K.~Nadezhin, and V.~S.~Pinaev, 
``Kinetic Equilibrium of $\beta$-Processes in Stellar Interiors,''
\emph{Soviet Astronomy} \textbf{10}, 970--978 (1967).

\bibitem{Arnett}
W.~D.~Arnett, and J.~W.~Truran, 
``Nucleosynthesis in Supernova Models. I. The Neutrino-Transport Model,''
\emph{ApJ} \textbf{160}, 959-970 (1970).

\bibitem{B03a}
A.~M.~Beloborodov, ``Nuclear Composition of Gamma-Ray Burst Fireballs,''
\emph{ApJ} \textbf{588}, 931--944 (2003).

\bibitem{Goodman}
J.~Goodman, A.~Dar, and S.~Nussinov,
``Neutrino Annihilation in Type II Supernovae,''
\emph{ApJ} \textbf{314}, L7--L10 (1987).

\bibitem{Eichler}
D.~Eichler, M.~Livio, T.~Piran, and D.~N.~Schramm,
``Nucleosynthesis, Neutrino Bursts and Gamma-rays from Coalescing 
Neutron Stars,''
\emph{Nature} \textbf{340}, 126--128 (1989).

\bibitem{Birkl}
R.~Birkl, M.~A.~Aloy, H.-Th.~Janka, and E.~M\"uller,
``Neutrino Pair Annihilation Near Accreting, Stellar-Mass Black Holes,''
\emph{A\&A} \textbf{463}, 51--67 (2007).

\bibitem{Sawyer}
R.~F.~Sawyer, ``Neutrino Transport in Accretion Disks,''
\emph{Phys. Rev. D} \textbf{68}, id. 063001 (2003).

\bibitem{Ramirez}
E.~Ramirez-Ruiz, and A.~Socrates,
``Supernovae and Gamma-Ray Bursts Powered by Hot Neutrino-Cooled Coronae,''
(arXiv:astro-ph/0504257)

\bibitem{Rossi2}
E.~M.~Rossi, P.~J.~Armitage, and T.~di~Matteo,
``Vertical Structure of Hyper-Accreting Disks and Consequences for 
Gamma-Ray Burst Outflows,''
\emph{A\&SS} \textbf{311}, 185--190 (2007).

\bibitem{Kawabata}
R.~Kawabata, S.~Mineshige, and N.~Kawanaka, N.
``Coronal Neutrino Emission in Hypercritical Accretion Flows,''
\emph{ApJ} \textbf{675}, 596--603 (2008).

\bibitem{BI}
A.~M.~Beloborodov, and A.~F.~Illarionov,
``Small-scale Inviscid Accretion Discs around Black Holes,''
\emph{MNRAS} \textbf{323}, 167--176 (2001).

\bibitem{Lee2}
W.~H.~Lee, and E.~Ramirez-Ruiz,
``Accretion Modes in Collapsars: Prospects for Gamma-Ray Burst Production,''
\emph{ApJ} \textbf{641}, 961--971 (2006).

\bibitem{Yakovlev}
D.~G.~Yakovlev, A.~D.~Kaminker, O.~Y.~Gnedin, and P.~Haensel,
``Neutrino Emission from Neutron Stars,''
\emph{Phys.~Rep.} \textbf{354}, 
1--155 (2001).

\bibitem{Igumen}
I.~V.~Igumenshchev, A.~F.~Illarionov, and M.~A.~Abramowicz, 
``Hard X-Ray-emitting Black Hole Fed by Accretion of Low Angular Momentum 
Matter,''
\emph{ApJ} \textbf{517}, L55--L58 (1999).

\bibitem{Lee1}
W. Lee, and E. Ramirez-Ruiz, ``The Progenitors of Short Gamma-Ray Bursts,''
\emph{New Journal of Physics} \textbf{9}, 
17--71 (2007). 

\bibitem{Metzger1}
B.~D.~Metzger, A.~L.~Piro, and E.~Quataert,
``Time-Dependent Models of Accretion Disks Formed from Compact Object 
Mergers,''
\emph{MNRAS}, submitted (arXiv:0805.4415)

\bibitem{Li}
L.-X.~Li, and B.~Paczy\'nski, ``Transient Events from Neutron Star Mergers,''
\emph{ApJ} \textbf{507}, L59--L62 (1998).

\bibitem{Lynden}
D.~Lynden-Bell, and J.~E.~ Pringle,
``The Evolution of Viscous Discs and the Origin of the Nebular Variables,''
\emph{MNRAS} \textbf{168}, 603--637 (1974).

\bibitem{Ogilvie}
G.~I.~Ogilvie, ``Time-Dependent Quasi-Spherical Accretion,''
\emph{MNRAS} \textbf{306}, L9--L13 (1999).

\bibitem{Narayan2}
R.~Narayan, and I.~Yi,
``Advection-Dominated Accretion: A Self-Similar Solution,''
\emph{ApJ} \textbf{428}, L13--L16 (1994).

\bibitem{Blandford}
R.~D.~Blandford, and M.~C.~ Begelman, 
``Two-dimensional Adiabatic Flows on to a Black Hole - I. Fluid Accretion,''
\emph{MNRAS} \textbf{349}, 68--86 (2004).

\bibitem{Derishev}
E.~V.~Derishev, V.~V.~Kocharovsky, and Vl.~V.~Kocharovsky,
``The Neutron Component in Fireballs of Gamma-Ray Bursts: 
Dynamics and Observable Imprints,''
\emph{ApJ} \textbf{521}, 640--64 (1999).

\bibitem{B03b}
A.~M.~Beloborodov, ``Neutron-fed Afterglows of Gamma-Ray Bursts,''
\emph{ApJ} \textbf{585}, L19--L22 (2003).

\bibitem{Rossi3}
E.~M.~Rossi, A.~M.~Beloborodov, and M.~J.~Rees,
``Neutron-Loaded Outflows in Gamma-Ray Bursts,''
\emph{MNRAS} \textbf{369}, 1797--1807 (2006).

\bibitem{Pruet}
J.~Pruet, T.~A.~Thompson, and R.~D.~Hoffman, 
``Nucleosynthesis in Outflows from the Inner Regions of Collapsars.''
\emph{ApJ} \textbf{606}, 1006--1018, (2004).

\bibitem{Barzilay}
Y.~Barzilay, and A.~Levinson, 
``Structure and Nuclear Composition of General Relativistic,
Magnetohydrodynamic Outflows from Neutrino-Cooled Disks,''
\emph{New Astr.} \textbf{16}, 6, 386--394 (2008).

\bibitem{Metzger2}
B.~D.~Metzger, T.~A.~Thompson, and E.~Quataert, 
``On the Conditions for Neutron-rich Gamma-Ray Burst Outflows,''
\emph{ApJ} \textbf{676}, 1130--1150 (2008).

\bibitem{Surman}
R.~Surman, G.~C.~McLaughlin, M.~Ruffert, H.-Th.~Janka, and W.~R.~Hix,
``r-Process Nucleosynthesis in Hot Accretion Disk Flows from Black 
Hole-Neutron Star Mergers,''
\emph{ApJ} \textbf{679}, L117--L120 (2008).

\end{thebibliography}



\end{document}